\RequirePackage{fix-cm}
\documentclass[a4paper,12pt]{article}
\usepackage[margin=1in]{geometry}
\usepackage{amsmath,amssymb,amsthm,mathtools,bm,booktabs,longtable,array,tabularx,graphicx}
\usepackage[round,authoryear]{natbib}
\usepackage[T1]{fontenc}
\usepackage{newtxtext,newtxmath,microtype}
\usepackage{setspace}
\usepackage{etoolbox}
\let\small\normalsize

\makeatletter
\let\submission@stretch\baselinestretch
\patchcmd{\@xfloat}{\def\baselinestretch{\setspace@singlespace}}
  {\let\baselinestretch\submission@stretch}{}{\PackageError{submission}{Float spacing patch failed}{}}
\patchcmd{\@footnotetext}{\def\baselinestretch{\setspace@singlespace}}
  {\let\baselinestretch\submission@stretch}{}{\PackageError{submission}{Footnote spacing patch failed}{}}
\patchcmd{\@mpfootnotetext}{\def\baselinestretch{\setspace@singlespace}}
  {\let\baselinestretch\submission@stretch}{}{\PackageError{submission}{Minipage footnote spacing patch failed}{}}
\makeatother
\usepackage{xr}
\usepackage[colorlinks=true,linkcolor=blue,citecolor=blue,urlcolor=blue]{hyperref}
\usepackage{caption}
\newcommand{\tablenotefont}{\normalsize\setstretch{1}}
\newcommand{\figurenotefont}{\tablenotefont}
\usepackage{enumitem}
\usepackage{placeins}
\usepackage{needspace}

\usepackage{authblk}

\setlist{nosep}
\allowdisplaybreaks
\newtheorem{assumption}{Assumption}
\newtheorem{theorem}{Theorem}
\newtheorem{proposition}{Proposition}

\theoremstyle{remark}\newtheorem{remark}{Remark}
\newtheorem{example}{Example}
\BeforeBeginEnvironment{assumption}{\Needspace{4\baselineskip}}
\BeforeBeginEnvironment{theorem}{\Needspace{6\baselineskip}}
\BeforeBeginEnvironment{proposition}{\Needspace{8\baselineskip}}
\BeforeBeginEnvironment{remark}{\Needspace{6\baselineskip}}
\BeforeBeginEnvironment{example}{\Needspace{5\baselineskip}}

\newcommand{\E}{\mathbb E}
\newcommand{\R}{\mathbb R}
\newcommand{\tr}{\operatorname{tr}}
\newcommand{\rank}{\operatorname{rank}}
\newcommand{\vecop}{\operatorname{vec}}

\newcommand{\norm}[1]{\left\lVert #1\right\rVert}
\newcommand{\ind}{\mathbf 1}

\AtBeginDocument{\setlength{\bibsep}{0pt}}

\title{Panel Local Projections with Missing Data\thanks{Yubo Tao acknowledges the financial support from the National Natural Science Foundation of China (Project No.~72303003) and a seed grant from the Asia-Pacific Academy of Economics and Management of the University of Macau.}}
\author[a]{Yubo Tao}
\affil[a]{\small\emph{Department of Economics, University of Macau, Macao SAR, China}}
\date{November 2026}
\IfFileExists{online_appendix.aux}{
  \externaldocument[][nocite]{online_appendix}
}{
  \AtBeginDocument{%
\newlabel{app:application_calibration}{{D.6}{101}{Calibration using the empirical data}{subsection.D.6}{}}
\newlabel{app:balanced}{{A.10}{52}{Balanced panels and the MSS benchmark}{subsection.A.10}{}}
\newlabel{app:balanced_simulation}{{C.3}{86}{Predictive panels with a common coefficient}{subsection.C.3}{}}
\newlabel{app:benchmark_score}{{A.10}{52}{Balanced panels and the MSS benchmark}{section*.3}{}}
\newlabel{app:conditional_proof}{{A.9}{49}{Observed-regressor inference and observation designs}{subsection.A.9}{}}
\newlabel{app:conditional_weights}{{A.9}{49}{Observed-regressor inference and observation designs}{subsection.A.9}{}}
\newlabel{app:dated_localization}{{B.6}{75}{Time-varying dependence weights}{subsection.B.6}{}}
\newlabel{app:feasible}{{A.4}{24}{Covariance estimation from observed pairs}{subsection.A.4}{}}
\newlabel{app:identification}{{A.2}{19}{Probability bounds and the factor fit}{section*.1}{}}
\newlabel{app:industry_data}{{D.2}{94}{Industry and peer data}{subsection.D.2}{}}
\newlabel{app:lmb}{{B.1}{62}{Fixed localization kernels}{section*.4}{}}
\newlabel{app:main_proof}{{A.5}{35}{Proof of the coefficient theorem}{subsection.A.5}{}}
\newlabel{app:monthly_screen}{{D.4}{96}{Monthly conditioning sensitivity}{subsection.D.4}{}}
\newlabel{app:pattern_proof}{{A.9}{51}{Observed-regressor inference and observation designs}{section*.2}{}}
\newlabel{app:regularization_discussion}{{A.1}{38}{Covariance regularization}{remark.A.1}{}}
\newlabel{app:sample_construction}{{D.1}{92}{Data construction and sample selection}{subsection.D.1}{}}
\newlabel{app:score_expansion}{{A.1}{2}{Moment weights and the factor geometry}{subsection.A.1}{}}
\newlabel{app:simulation_design}{{C.1}{78}{Data-generating processes and observation schedules}{subsection.C.1}{}}
\newlabel{app:target_comparison}{{D.5}{99}{Sample periods and the projection coefficient}{subsection.D.5}{}}
\newlabel{cor:covariance_adjustment}{{A.1}{32}{Covariance refinement under positive overlap}{corollary.A.1}{}}
\newlabel{cor:window_lmb}{{A.2}{58}{Localized multiplier bootstrap for prespecified dates}{corollary.A.2}{}}
\newlabel{eq:I12}{{A.61}{39}{Joint-score covariance and standard errors}{equation.A.61}{}}
\newlabel{eq:U22}{{A.7}{3}{Moment weights and the factor geometry}{equation.A.7}{}}
\newlabel{eq:feasibletrace}{{A.9}{3}{Moment weights and the factor geometry}{equation.A.9}{}}
\newlabel{fig:industry_simultaneous}{{D.1}{97}{Simultaneous confidence bands for monthly investment responses}{figure.caption.12}{}}
\newlabel{fig:localized_simulation_coverage}{{C.1}{90}{Pointwise coverage across monthly horizons}{figure.caption.7}{}}
\newlabel{fig:quarterly_common_calendar}{{D.3}{100}{Quarterly responses for firms with records covering the full period}{figure.caption.14}{}}
\newlabel{lem:P2}{{A.3}{8}{Expansion of the fitted common component}{lemma.A.3}{}}
\newlabel{lem:lmb_counterexample}{{B.4}{74}{Fitted row normal equations}{lemma.B.4}{}}
\newlabel{lem:lmb_dependence}{{B.1}{63}{Score dependence}{lemma.B.1}{}}
\newlabel{lem:lmb_score_error}{{B.2}{65}{Estimated-score error}{lemma.B.2}{}}
\newlabel{lem:rare}{{A.9}{27}{Rare-overlap functional bounds}{lemma.A.9}{}}
\newlabel{prop:known_windows}{{A.2}{55}{Prespecified estimation dates}{proposition.A.2}{}}
\newlabel{prop:lmb_bands}{{B.2}{72}{Fixed-horizon simultaneous bands}{proposition.B.2}{}}
\newlabel{prop:peer_features}{{B.3}{77}{Mutual-peer features}{proposition.B.3}{}}
\newlabel{subsec:thresholded_inference}{{A.7}{38}{Thresholded score covariance}{subsection.A.7}{}}
\newlabel{tab:general_coverage}{{C.3}{91}{Missing-regressor inference with asymmetric innovations and internal gaps}{table.caption.8}{}}
\newlabel{tab:gpr_missingness}{{D.1}{92}{Quarterly variable availability, 912 firms over 1995--2025}{table.caption.9}{}}
\newlabel{tab:gpr_nested}{{D.2}{93}{Investment responses at the two-quarter horizon}{table.caption.10}{}}
\newlabel{tab:localized_patterns}{{C.2}{89}{Localized multiplier bootstrap coverage by observation pattern}{table.caption.6}{}}
  }
}
\makeatletter
\renewcommand{\@maketitle}{%
  \newpage
  \begin{center}%
    {\LARGE\@title\par}%
    \vspace{0.4em}%
    {\large\@author\par}%
    \vspace{0.3em}%
    {\large\@date\par}%
  \end{center}%
  \par\vspace{0.3em}%
}
\makeatother
\renewenvironment{abstract}{%
  \begin{center}\bfseries\abstractname\end{center}%
  \noindent\ignorespaces
}{\par}
\begin{document}
\maketitle
\begin{abstract}
A panel local projection can remain identified even when missing data leave no complete regression case. We recover the projection over specified dates by combining factor-based conditional means with idiosyncratic covariances from observed pairs. Under strong factors in the auxiliary panel, sufficient factor variation at observed dates and enough observations of the required pairs, a second-order expansion yields bias-corrected moments and a joint score for inference. The localized multiplier bootstrap provides pointwise confidence intervals and simultaneous bands under additional dependence restrictions. We revisit the investment response to geopolitical risk studied by \citet{CI2022} with a larger quarterly firm sample and extend the analysis to monthly projections of trailing-three-month investment intensity. In the monthly analysis, the six-month response difference between high- and low-exposure firms becomes more negative as firms with incomplete histories enter the sample.
\end{abstract}

\clearpage
\section{Introduction}
Panel local projections trace responses at successive horizons, each of which may require a different combination of current variables, lagged controls and future outcomes \citep{Jorda2005}. With incomplete histories, the units and dates available for estimation can change along the response path. If reporting windows cover different economic conditions, the resulting coefficients can describe different projections. Even a long history for each variable may leave few dates on which all required variables are observed together. We develop a method that uses these incomplete histories to estimate a projection for a given population over specified dates.

The distinction between complete regression cases and projection moments is central to the problem. A complete case requires the outcome and the entire regressor vector at the same base date. Each entry of the normal equations, however, involves only a pair of variables, and different pairs can be observed on different dates. If every third date is missing, a specification requiring three adjacent dates can lose every complete case while retaining repeated observations of each required pair. The projection can therefore remain identified. What matters is whether the observed pairs contain enough information to recover the moments for the intended period.

Fully observed auxiliary series provide information about the common factors entering the outcome and regressors. We use them to estimate conditional means at missing entries and combine these means with idiosyncratic covariances estimated from observed residual pairs. Conditional on the common component and observation pattern, these within-unit covariances are time invariant. As \citet{CBN2023} explain, conditional-mean imputation alone omits the idiosyncratic variation needed for covariance estimation. Restoring that variation in the projection moments gives the moment-restored local projection estimator, MRLP.

Demeaning and estimating the missing contributions introduce further bias. The leading terms have order $N^{-1}+T^{-1}$, which is also the standard-error order when $N$ and $T$ are comparable. We derive a second-order expansion and construct the analytical correction MRLP-BC. Its leading score contains linear estimation errors and centered quadratic errors from restored residual products. Third innovation moments can make these terms correlated. The expansion carries both terms through demeaning and inversion of the regressor second-moment matrix. With complete observations, the factor-estimation correction vanishes exactly and the demeaning correction remains.

Different pairwise observations are needed to identify the projection and estimate its bias correction. Every idiosyncratic covariance entering a projection moment must be based on enough joint observations. Other covariance pairs, used for the bias correction, may have few or no observations. We omit estimates based on too few pairs and bound their aggregate contribution on the observed entries. For inference, a localized multiplier bootstrap applies correlated weights to the complete estimated score. It gives pointwise intervals and simultaneous bands over a fixed set of horizons when the weights preserve the covariances that contribute to the limiting variance.

These questions arise within the broader literature on local projections and dynamic panels. Local projection theory relates response coefficients to vector autoregressions \citep{PMW2021} and develops inference for their estimates \citep{MOPM2021,JT2025}. The instrumental-variable and generalized method of moments (GMM) estimators of \citet{AB1991}, \citet{ABo1995} and \citet{BB1998} identify common coefficients from available transformed equations and, under additional initial-condition restrictions, equations in levels. Unbalanced panels are usable when their observed equations and instruments satisfy the required orthogonality conditions. A projection over a specified period can also depend on the temporal weights given to common economic conditions. Recovering those full-period moments is a distinct task.

The within transformation creates the bias studied by \citet{Nickell1981} and corrected through approximations such as \citet{Kiviet1995}. \citet{Bruno2005} extends this approach to unbalanced panels under strictly exogenous selection, including gaps in the histories. For large $N$ and $T$, \citet{HK2002}, \citet{HN2004} and \citet{FVW2016} develop analytical and jackknife corrections for fixed-effects models. In panel local projections, \citet{Mei2026}, denoted MSS, use a split-panel jackknife related to \citet{DJ2015}. Their extension to unbalanced panels permits consecutive histories of unequal length when the full and half histories identify the same slope. It requires no model for missing values. Our method imposes a factor structure to recover the intended moments when the coefficients over the full and observed periods differ. The simulations compare the two corrections directly where the histories identify the same coefficient.

Observation patterns and selection restrictions must also be distinguished. \citet{Wooldridge1995} develops selection corrections under conditional mean restrictions, while \citet{Kyriazidou2001} and \citet{SemykinaWooldridge2013} address selection in dynamic panels. Inverse probability weighting treats selection and attrition under appropriate ignorability conditions \citep{Wooldridge2002}. \citet{CSUnbalanced2026} study two-way fixed-effects inference with deterministic or stochastic selection, including feedback from past outcomes under their conditional mean restriction. Here, conditioning on the observation pattern preserves identification only when the factor and idiosyncratic moment restrictions continue to hold. The admissible observation patterns are broad, but do not include arbitrary endogenous attrition.

Factor models supply information about the missing series \citep{Bai2003,BN2006}. In interactive fixed-effects regressions, factors enter the regression error and must be accounted for to estimate its slope \citep{Bai2009,MW2017}, including in the unbalanced analyses of \citet{SWW2026} and \citet{CS2026}. Our factors describe the outcome and regressor series themselves. Their products contribute to the projection moments, so including them as additional regression controls would generally change the target. The construction instead uses completion methods \citep{BN2021,XP2023,SW2025} to recover that contribution. The case of sufficient auxiliary information in \citet[Section~6.1]{DPX2024} provides the relevant starting point: complete auxiliary series span every factor needed for the target panel. We use these series to initialize the factors and then estimate the model jointly with the observed target data.

Following \citet{SW2025}, observation indicators describe the missingness pattern, whose number of distinct histories may grow with the panel. Random missingness, leading and trailing blocks, internal gaps, ragged edges and mixed-frequency reporting can satisfy the conditions. Each target row must retain enough factor variation to identify its loading, and every pair entering the projection must have enough observations. Reporting frequency makes the latter restriction especially clear. Quarterly observations may identify a conditional mean on a monthly grid without identifying an unrestricted idiosyncratic lag-one covariance. A conditional projection with fully observed regressors avoids that missing covariance. We develop this case separately and allow prespecified unit-specific projection windows.

Recovery from different observed moments also connects to missing covariates and auxiliary-data methods \citep{RRZ1994,CHT2008,Graham2011} and to unequal moment histories in GMM \citep{LW2013}. \citet{DP2026} provide the closest componentwise imputation comparison. They correct moments using observed covariate pairs, and their panel extension permits temporal dependence and time-varying marginal distributions. Their unit-level cross-fitting argument treats trajectories as independent. Our conditional large-$N,T$ expansion instead carries the effects of the joint factor fit and covariance restoration through a score allowing weak cross-unit dependence. Related low-rank inference studies entries, functionals and treatment effects \citep{Athey2021,CHLZ2023,CKL2024,CY2024,CKL2026}. The projection additionally requires demeaning and inversion of estimated moments. Pattern-specific conditional-expectation methods \citep{RAY2026} use different restrictions. Our results concern the stated factor model and supply no general efficiency ordering over these approaches.

The Monte Carlo simulations distinguish moment recovery from bias correction. DGPs with changing factor variation examine recovery of the full-period projection. A common-coefficient design compares MSS and MRLP-BC on consecutive histories using $N=200$ or $1000$ and $T=360$ or $120$. A conditional-projection design studies monthly horizons under quarterly reporting. The localized multiplier bootstrap reduces conservative coverage in its principal mixed-frequency comparisons, while undercoverage remains in a short-panel block case. All main designs retain strong auxiliary factor information.

We revisit the investment question in \citet{CI2022}: whether investment falls more in industries with greater exposure to geopolitical risk. Incomplete quarterly histories allow us to include more firms, and monthly CRSP data extend the analysis to trailing-three-month investment intensity between fiscal quarter ends. The monthly estimates use a separate sample period for each firm, bounded by its first and last quarterly records. The six-month response difference between high- and low-exposure firms becomes more negative as firms with incomplete histories enter the sample. Estimates over a common period and a restriction based on observed factor variation show how the results change with the sample and estimation period. These comparisons describe different projections and do not test equality of the population responses.

Section~\ref{sec:model} presents the model and identification conditions. Section~\ref{sec:theory} develops estimation and inference. Section~\ref{sec:simulations} reports the Monte Carlo simulations. Section~\ref{sec:application} examines the investment response to geopolitical risk. Section~\ref{sec:conclusions} concludes.

\section{Panel Local Projections with Missing Observations}\label{sec:model}
\subsection{The local projection and factor model}\label{subsec:target}
For a fixed horizon $h$, consider the panel local projection
\begin{equation}\label{eq:lp_model}
 y_{i,t+h}=\alpha_{ih}+x_{it}'\beta_{h,NT}+u_{it,h}.
\end{equation}
Here $y_{i,t+h}$ is the future outcome, $x_{it}\in\R^p$ contains a fixed set of current variables and lagged controls, and $\alpha_{ih}$ is a unit effect. The coefficient $\beta_{h,NT}$ summarizes the within-unit predictive relation over a specified full period. We define it through an expected least-squares criterion, with $u_{it,h}$ its projection error. This formulation gives the usual panel local projection \citep{Jorda2005,Mei2026} while separating the period defining its target from the dates available for estimation. A causal interpretation requires additional identifying restrictions.

Units are indexed by $i=1,\ldots,N$, and the underlying series span dates $t=1,\ldots,T$. Let $\mathcal T_h$ be the common set of regression dates for which the specified leads and lags lie within this period, with $T_h=|\mathcal T_h|=T-O(1)$. The regression date refers to the current regressors. Leads and lags are measured relative to that date. Write $n=NT_h$ for the number of unit--time pairs. The set $\mathcal T_h$ includes dates with missing outcomes or regressors. Our objective is to estimate the projection over this full period from the available observations.

The full-period normal equations cannot be computed directly when their component variables are missing. We recover them by modeling the underlying outcome and regressor series jointly with auxiliaries. Following the observation-indicator formulation of \citet{SW2025}, let, for variable $a=1,\ldots,q$,
\begin{equation}\label{eq:scalar_model}
 z_{it,a}=\eta_{ia}+\lambda_{ia}'f_t+e_{it,a}.
\end{equation}
The vector $f_t\in\R^r$ contains the common factors, $\lambda_{ia}\in\R^r$ contains the loadings for variable $a$ of unit $i$, $\eta_{ia}$ is its individual fixed effect, and $e_{it,a}$ is its idiosyncratic error. These errors differ from the projection error $u_{it,h}$ in \eqref{eq:lp_model}. The fixed effects may be correlated with the loadings and the missing data pattern. The number of variables $q$ and the number of factors $r$ are fixed.

Let $R_{it,a}=\ind\{z_{it,a}\text{ is observed}\}$. Stack the variables and units in the $qN\times T$ matrices $Z,C,E$ and $R$, where $C_{it,a}=\eta_{ia}+\lambda_{ia}'f_t$. In matrix form,
\begin{equation}\label{eq:model}
 Z=C+E,\qquad C=\eta\ind_T'+L,\qquad L=\Lambda F',\qquad
 F'\ind_T=0,\quad F'F/T=I_r,\quad\rank(L)=r.
\end{equation}
We refer to the sum of the individual fixed effects and the time-varying factor component as the common component. Here $\Lambda\in\R^{qN\times r}$ stacks the loadings, $F\in\R^{T\times r}$ stacks the factors, and $\ind_T$ is the $T$-vector of ones. The centering restriction separates the individual fixed effects from the time-varying factor component, and the second restriction normalizes the factor scale. The observed data are $(R,R\circ Z)$, where $\circ$ denotes entrywise multiplication and zeros at missing entries are placeholders. For matrices, $\|\cdot\|$ and $\|\cdot\|_F$ denote spectral and Frobenius norms, respectively.

\citet{SWW2026} use observation indicators in unbalanced regressions with interactive fixed effects. In \eqref{eq:scalar_model}, the factors instead describe the underlying series from which the projection moments are formed. Their products remain in the target. Absorbing those factors as additional controls would remove part of the variation that the present coefficient is intended to summarize.

We condition on $(C,R)$. Under the conditional mean restriction in Assumption~\ref{ass:moments}, $C$ is the panel of conditional means. The population projection coefficients solve
\begin{equation}\label{eq:feobjective}
 (\alpha_{1h},\ldots,\alpha_{Nh},\beta_{h,NT})
 =\arg\min_{\alpha_1,\ldots,\alpha_N,b}
 \frac1n\sum_{i,t\in\mathcal T_h}
 \E[(y_{i,t+h}-\alpha_i-x_{it}'b)^2\mid C,R].
\end{equation}
Let $c_{x,it}=\E[x_{it}\mid C,R]$ and $c_{y,i,t+h}=\E[y_{i,t+h}\mid C,R]$. Their within-unit averages and the projection fixed effects are
\[
 \bar c_{x,i}=T_h^{-1}\sum_{t\in\mathcal T_h}c_{x,it},\qquad
 \bar c_{y,i}=T_h^{-1}\sum_{t\in\mathcal T_h}c_{y,i,t+h},\qquad
 \alpha_{ih}=\bar c_{y,i}-\bar c_{x,i}'\beta_{h,NT}.
\]
Decompose the projection variables as $x_{it}=c_{x,it}+e_{x,it}$ and $y_{i,t+h}=c_{y,i,t+h}+e_{y,i,t+h}$, where the idiosyncratic terms inherit the same leads and lags as the corresponding variables. The slope normal equations are
\begin{equation}\label{eq:target}
\begin{aligned}
 Q_{h,NT}&=\frac1n\sum_{i,t\in\mathcal T_h}
 \{(c_{x,it}-\bar c_{x,i})(c_{x,it}-\bar c_{x,i})'
       +\E[e_{x,it}e_{x,it}'\mid C,R]\},\\
 q_{h,NT}&=\frac1n\sum_{i,t\in\mathcal T_h}
 \{(c_{x,it}-\bar c_{x,i})(c_{y,i,t+h}-\bar c_{y,i})
       +\E[e_{x,it}e_{y,i,t+h}\mid C,R]\},\\
 \beta_{h,NT}&=Q_{h,NT}^{-1}q_{h,NT}.
\end{aligned}
\end{equation}
Each moment has two parts. Centered conditional means retain the common-component variation over the intended projection period, and idiosyncratic covariances retain the remaining variation. Mean imputation recovers only the first part. For a fixed horizon write $Q_{NT}=Q_{h,NT}$, $q_{NT}=q_{h,NT}$ and $\beta_0=\beta_{h,NT}$. Fixed finite sums, including debt accumulation and cumulative growth, enter through the corresponding sums of pairwise moments.

For pair type $k$, let $(a_k,b_k)$ be the variables and $(\ell_{1k},\ell_{2k})$ their offsets. Set $\mathcal T_k=\mathcal T_h$, $T_k=T_h$, $n_k=NT_k$, and
\[
 \bar C_{ik}^{(s)}=T_k^{-1}\sum_{t\in\mathcal T_k}
 C_{i,t+\ell_{sk},a_{sk}},\qquad (a_{1k},a_{2k})=(a_k,b_k).
\]
Every entry of $Q_{NT}$ and $q_{NT}$ is a linear combination of
\begin{equation}\label{eq:moments}
 \mu_k=F_k(C)+\frac1N\sum_i\Gamma_{ik},\qquad
 F_k(C)=\frac1{n_k}\sum_{i,t\in\mathcal T_k}
 (C_{i,t+\ell_{1k},a_k}-\bar C_{ik}^{(1)})
 (C_{i,t+\ell_{2k},b_k}-\bar C_{ik}^{(2)}).
\end{equation}
Here $\Gamma_{ik}=\E[e_{i,t+\ell_{1k},a_k}e_{i,t+\ell_{2k},b_k}\mid C,R]$ is time-invariant within unit. Write $\mathcal I_k$ for these entry pairs and $\mathcal I_{ik}$ for the pairs of unit $i$. The number of pair types is fixed.

The fixed effects in \eqref{eq:feobjective} minimize an expected criterion. Sample demeaning in a balanced panel also subtracts the random means $\bar e_{x,i}$ and $\bar e_{y,i}$. Its expected within-estimator score at $\beta_0$ equals
\begin{equation}\label{eq:withinbias}
 -\frac1N\sum_i\E[\bar e_{x,i}(\bar e_{y,i}-\bar e_{x,i}'\beta_0)\mid C,R].
\end{equation}
Under short memory, this term is generally of order $T^{-1}$ and need not vanish in dynamic projections. Our estimator retains observed values and fills missing entries with estimated conditional means. Its correction restores both the missing idiosyncratic covariance contribution and the mean-product term removed by sample demeaning. A separate second-order term accounts for estimation of the factor model.

\phantomsection\label{app:examples_patterns}
The following examples show how familiar projections fit this moment representation. Write $\mu_{ab}(u,v)$ for the moment in \eqref{eq:moments} involving variables $a,b$ at offsets $u,v$. All moments in a given projection use the same admissible regression dates.

\begin{example}[A predictive regression with periodic missing observations]\label{ex:intermittent}
Let $d_{it}$ be an economic predictor and consider
\[
 y_{i,t+1}=\alpha_{i1}+\beta_{1,NT}^{(1)}d_{it}
                     +\beta_{1,NT}^{(2)}d_{i,t-1}+u_{it,1}.
\]
The superscripts distinguish the two coordinates of the horizon-one slope $\beta_{1,NT}$. Here $x_{it}=(d_{it},d_{i,t-1})'$, and the normal equations in \eqref{eq:target} use
\[
 Q_{1,NT}=\begin{pmatrix}
 \mu_{dd}(0,0)&\mu_{dd}(0,-1)\\
 \mu_{dd}(0,-1)&\mu_{dd}(-1,-1)
 \end{pmatrix},\qquad
 q_{1,NT}=\begin{pmatrix}\mu_{dy}(0,1)\\\mu_{dy}(-1,1)\end{pmatrix}.
\]
Suppose the predictor and outcome of unit $i$ are missing at dates $t\bmod3=g_i$ and observed otherwise, for some $g_i\in\{0,1,2\}$. The regression requires observations at $t-1,t,t+1$, so every case has a missing entry. Yet each squared regressor is observed on two-thirds of the dates, and each cross-product above on one-third, up to boundary terms. The normal equations therefore require less joint observation than the regression itself.

Fully observed auxiliary series supply information about the factor paths in \eqref{eq:scalar_model}. Conditional on the means, the observed pairs identify each unit's idiosyncratic covariance under the restrictions below. Combining the two components recovers the full-period normal equations. This example motivates the periodic-missingness design in Section~\ref{sec:simulations}. Identification requires strong auxiliary factor information, sufficient factor variation on observed dates and time-invariant within-unit idiosyncratic covariances.
\end{example}

\begin{example}[Cumulative growth with unequal consecutive debt histories]\label{ex:cumulative}
Let $d_{it}$ be the annual change in a country's debt-to-GDP ratio and $y_{it}$ its annual GDP growth. For fixed positive integers $m$ and $h$, define past debt accumulation and subsequent cumulative growth by
\[
 A^{(m)}_{it}=\sum_{s=1}^m d_{i,t-s},\qquad
 G^{(h)}_{it}=\sum_{v=1}^h y_{i,t+v}.
\]
With a vector of controls $w_{it}$ of fixed dimension, the projection is
\[
 G^{(h)}_{it}=\alpha_{ih}+\theta_{h,NT}A^{(m)}_{it}
                         +w_{it}'\gamma_{h,NT}+u_{it,h}.
\]
It has the form \eqref{eq:lp_model} with outcome $G^{(h)}_{it}$, regressor vector $x_{it}=(A^{(m)}_{it},w_{it}')'$ and slope $(\theta_{h,NT},\gamma_{h,NT}')'$. The regressor second moment for debt accumulation and its cross-moment with growth are, respectively,
\[
 \sum_{s=1}^m\sum_{j=1}^m\mu_{dd}(-s,-j),\qquad
 \sum_{s=1}^m\sum_{v=1}^h\mu_{dy}(-s,v).
\]
Thus cumulative variables require a fixed collection of annual pairwise moments. The controls are handled by the same expansion when they contain fixed lags or finite sums of the underlying series.

Suppose GDP growth and auxiliary consumption growth are fully observed, while debt changes are observed only from a country-specific date $t_i$ onward. Complete-case regression represents countries over different periods. Moment restoration fixes a common regression period, estimates conditional means over that period and uses fitted residual pairs from each country's observed dates. The coefficient $\theta_{h,NT}$ describes the partial predictive association for that common period. A causal interpretation requires further identifying restrictions. Start dates may differ across countries. Identification requires positive pair overlap, sufficient observed factor variation and the stated moment restrictions. This specification illustrates the debt and growth projections studied by \citet{MSV2017}.
\end{example}

\subsection{Identification from auxiliary series and observed pairs}\label{subsec:moment_conditions}\label{subsec:completion}

Let $\mathcal A$ be the set of rows corresponding to $q_0\ge1$ auxiliary variables observed at every date, where $q_0<q$. The auxiliary panel must contain enough information to recover every factor entering the target variables. This corresponds to the case with sufficient auxiliary information in \citet{DPX2024}. The loading condition below states this requirement explicitly.

Let $S_j=\{t:R_{jt}=1\}$ and $x_t^f=(1,f_t')'$. The observed
sets may differ across all rows. We impose separate conditions on factor information and on the pairs
needed for the projection moments.

\begin{assumption}[Auxiliary information and observed factor variation]\label{ass:calendar}
For fixed $q,q_0,r$, $0<c_0<C_0<\infty$ and $B_0<\infty$,
\begin{enumerate}[label=(\roman*)]
\item $\sup_j\|\lambda_j\|+\sup_t\|f_t\|\le C_0$,
$F'F/T=I_r$, $F'\ind_T=0$, and
$\|\Lambda F'\|\le B_0\sqrt{NT}$.
\item $R_{jt}=1$ $(j\in\mathcal A)$ and
$N^{-1}\Lambda_{\mathcal A}'\Lambda_{\mathcal A}\succeq c_0I_r$.
\item $c_0I_{r+1}\preceq T^{-1}\sum_{t\in S_j}x_t^fx_t^{f\prime}
       \preceq C_0I_{r+1}\quad(j=1,\ldots,qN)$.
\end{enumerate}
\end{assumption}

\begin{assumption}[Pairwise observation frequencies]\label{ass:pairs}
For every pair type $k$ entering $Q_{NT}$ or $q_{NT}$,
\begin{equation}\label{eq:obs_gram}
 \inf_{i,k}p_{ik}\ge c_0>0,\qquad
 p_{ik}=T_h^{-1}\sum_{(u,v)\in\mathcal I_{ik}}R_uR_v.
\end{equation}
\end{assumption}

Assumption~\ref{ass:calendar}(i) centers and normalizes the factors, separating the time-varying component from the unrestricted fixed effects. Bounded factor and loading rows control the factor contribution to the projection moments, while the spectral bound sets the scale of the constraint in \eqref{eq:affinefit}.

Assumption~\ref{ass:calendar}(ii) provides factor information at every date through the complete auxiliary block. Its loading Gram bound requires every relevant factor to be strong in that block. This is the case with sufficient auxiliary information of \citet{DPX2024}, whose more general analysis studies weighting panels with different factor information. Target observations also enter the joint fit.

Assumption~\ref{ass:calendar}(iii) requires informative factor variation on each row's observed dates to identify its loading and fixed effect. Observation counts alone do not ensure this. We maintain strong factors \citep{BN2021,CBN2023,SW2025} and require the observed-Gram eigenvalues to remain bounded away from zero. A small fixed positive lower bound can still give an inaccurate finite-sample approximation.

Assumption~\ref{ass:pairs} requires observed idiosyncratic error pairs for each unit, rather than complete regression cases. Its positive lower bound controls the inverse-overlap weights $R_uR_v/p_{ik}$. Allowing required fractions to vanish would need a separate analysis of these weights and rates, even when common means remain recoverable. Nuisance row-pair overlaps may be zero or one. Section~\ref{subsec:conditional} removes the missing regressor--outcome moment requirement for fully observed regressors.

\begin{remark}[Relation to missing-data patterns in the literature]\label{rem:missing_patterns}
We use \emph{heterogeneous random missingness}, \emph{block missingness}, \emph{mixed-frequency observations} and \emph{ragged edges} as in \citet{SW2025,DPX2024}. Consecutive histories may have leading or trailing missing blocks. \emph{Internal gaps} lie between the first and last observed values, while \emph{intermittent missingness} involves repeated disappearances and returns. Ragged edges describe unequal availability near a panel boundary, including release delays. These descriptions can overlap.

Staggered treatment adoption in \citet{DPX2024} produces a trailing block of unobserved untreated potential outcomes. The broader staggered setup in \citet{SW2025} permits unrestricted entries outside informative complete rows and an initial common observation window, including later internal gaps. Intermittency alone establishes neither this structure nor the required information. Rearranging mixed-frequency observations into blocks does not change the local projection's original leads and lags.

Our complete auxiliaries can identify target means even at dates with no target observations, provided each target row retains informative factor variation. Projection recovery also requires observed idiosyncratic pairs, so mean recovery alone need not identify a projection involving an unobserved lagged outcome. Our conditions do not nest either paper's entire model class, including cases without sufficient complete auxiliaries or with vanishing observation fractions.

\end{remark}

The following observation patterns give sufficient conditions for
Assumption~\ref{ass:calendar}. A general projection must also satisfy
Assumption~\ref{ass:pairs}.
\begin{proposition}[Observation patterns]\label{prop:patterns}
Maintain bounded $x_t^f$, an informative complete auxiliary block and
$T^{-1}\sum_tx_t^fx_t^{f\prime}\succeq cI$. The observed Gram restriction
holds in each of the following cases.
\begin{enumerate}[label=(\roman*)]
\item $I_j\subseteq S_j$ and
$T^{-1}\sum_{t\in I_j}x_t^fx_t^{f\prime}\succeq cI$ for every $j$.
\item For fixed $m$, $P_a=\{t:t\equiv a\pmod m\}$,
$T^{-1}\sum_{t\in P_a}x_t^fx_t^{f\prime}\succeq c_1I$ for every reporting position used,
$S_j\supseteq P_{a_j}\setminus D_j$, and
$T^{-1}\sum_{t\in D_j}\|x_t^f\|^2\le c_1-c$.
\item Conditional on $C$, $\{R_{jt}:t\le T\}$ are independent Bernoulli
with $\inf_{j,t}\Pr(R_{jt}=1\mid C)\ge p_0>0$. The conclusion holds
jointly over $qN\asymp T$ target rows with probability tending to one.
\end{enumerate}
\end{proposition}

Proposition~\ref{prop:patterns}(i) covers leading and trailing blocks, ragged edges and internal deletions when the retained observations remain informative. Entry and exit refer to reporting boundaries. The full-period projection presumes that its variables are economically defined throughout the period. Proposition~\ref{prop:patterns}(ii) covers mixed-frequency reporting and gaps within a reporting schedule. Proposition~\ref{prop:patterns}(iii) allows heterogeneous random observation, subject to the conditional disturbance restrictions. The proof is in Online Appendix~\ref*{app:pattern_proof}.

In Example~\ref{ex:intermittent}, $S_i=\{t:t\bmod K\ne g_i\}$ with
$K=3$. More generally, $K\ge3$ gives
$p_{ik}\ge(K-2)/K+O(T^{-1})$, since two entries exclude at most two
residue classes. By contrast, observing only one date in every three can
give zero overlap at adjacent dates. Such an observation pattern can identify the common
mean under Proposition~\ref{prop:patterns}(ii) without identifying an unrestricted lag-one
projection.

Adding a row constant changes only its fitted fixed effect, leaving the fitting criterion, residuals and within moments unchanged.

\begin{assumption}[Idiosyncratic moments and regressor variation]\label{ass:moments}
Conditional on $(C,R)$,
\[
 \E[e_{it,a}]=0,\qquad
 \E[e_{it,a}e_{i,t-\ell,b}]=\Gamma_{i,ab}(\ell).
\]
$Q_{NT}\to Q\succ0$ and $\norm{\beta_0}$ is bounded.
\end{assumption}

Assumption~\ref{ass:moments} requires time-invariant within-unit idiosyncratic covariances conditional on the common component and observation pattern. Covariances may differ across units. Their time invariance connects observed pairs to the full period, while the common component can change over time. The observed series therefore need not be stationary. As in \citet{CBN2023}, means alone do not recover the required second moments.

Conditioning on $(C,R)$ permits an observation pattern related to factors, loadings or unit-specific variances only when the stated conditional moments continue to hold. For example, observation patterns generated using the common component can satisfy the restriction when the idiosyncratic disturbances are generated independently of those draws. Selection on a contemporaneous unobserved disturbance generally changes both its conditional mean and its observed-pair covariance. A random or intermittent pattern need not satisfy this restriction. This is the identifying restriction used here. It differs from a missing-at-random condition given observed covariates, as considered in missing-data efficiency analyses \citep{CHT2008,Graham2011}.

The positive-definite limit of $Q_{NT}$ and bounded $\beta_0$ control inversion and the moment weights. Conditioning fixes the common component but retains its products in \eqref{eq:target}. Factor information recovers variation rather than partialling it out as in interactive-effects regressions \citep{Bai2009,SWW2026}. These conditions supply no causal interpretation.

The following proposition combines identification from auxiliary conditional means and observed pairs.

\begin{proposition}[Identification of the projection]\label{prop:identification}
Let $n=NT_h$, $m=T_h$, and $k=(a,b,\ell_1,\ell_2)$. Under
Assumptions~\ref{ass:calendar}--\ref{ass:moments}, the observed-data law identifies $C$ and
\begin{equation}\label{eq:identified_moment}
\begin{aligned}
 \Gamma_{ik}
 &=\frac1{mp_{ik}}\sum_{(u,v)\in\mathcal I_{ik}}R_uR_v
       \{\E[Z_uZ_v\mid C,R]-C_uC_v\},\\
 \mu_k&=F_k(C)+\frac1N\sum_i\Gamma_{ik},\qquad
 \beta_0=Q_{NT}^{-1}q_{NT}\quad\text{if }\lambda_{\min}(Q_{NT})>0.
\end{aligned}
\end{equation}
Here $F_k(C)$ is the full-period within-unit product in \eqref{eq:moments}.
\end{proposition}

Proposition~\ref{prop:identification} identifies the coefficient even when no regression case is complete.\footnote{For the proof, see Online Appendix~\ref*{app:identification}.} The common-component term in \eqref{eq:identified_moment} uses the full projection period for averaging and centering. The covariance term uses each unit's available pairs with their own observation fractions, allowing different moments to use different histories \citep{LW2013}. Observed common-component products need not represent the full period.

Pairwise recovery is also a feature of the componentwise correction in \citet{DP2026}. The proposition specifies the factor and covariance restrictions under which the required pair moments are identified here. Joint estimation of these components on the same panel introduces the bias and variance terms considered next.

\phantomsection\label{app:centering_details}
Centering over observed dates can remove factor variation. If values $(-1,0,1)$ repeat every three dates and a row is observed at the first two positions, its observed factor mean is $-1/2$, while its full-period mean is zero. Joint estimation of fixed effects and loadings separates these terms and permits centering over the intended period. The period fixes the projection, while the observation pattern determines its recovery under the factor and covariance restrictions.

\section{Estimation and Inference}\label{sec:theory}
Estimation follows the two-part structure of the normal equations. We retain observed values because they contain both common and idiosyncratic variation. Fitted means supply the missing common component, and observed residual pairs supply the covariance contribution that their products omit. A further correction is needed because demeaning and estimation of these components can affect inference at first order.

\subsection{The three-step estimator}\label{sec:estimator}\label{subsec:pair_estimator}\label{subsec:estimator_summary}\label{sec:comparison}
Fix the projection horizon, regressors, and factor rank $r$. We compute the bias-corrected, moment-restored local projection (MRLP-BC) in the following three steps.

\begin{enumerate}[label={},leftmargin=0pt,labelsep=0pt,itemindent=0pt,itemsep=\baselineskip,topsep=0.5\baselineskip]
\item \textbf{Step 1. Estimate the common component.}
Estimate $\widehat C$ from the observed entries to recover the conditional means at missing dates. Following the observed-entry least-squares formulation in \citet[Section~2.2]{SW2025} and \citet[Section~3]{SWW2026}, let $\mathcal P_RX=R\circ X$ and minimize
\begin{equation}\label{eq:affinefit}
 \mathcal J(a\ind_T'+L_D)=\tfrac12\|R\circ(Z-a\ind_T'-L_D)\|_F^2,
 \qquad L_D\ind_T=0,\quad\rank(L_D)\le r,\quad\|L_D\|\le B\sqrt{NT},
\end{equation}
where $B>B_0$ and $B_0\sqrt{NT}$ bounds the true time-varying component. Write the fitted panel as $\widehat C=\widehat\eta\ind_T'+\widehat\Lambda\widehat F'$. The fit must meet the objective and projected-gradient requirements stated below. Global convergence of alternating least squares is not presumed.

\phantomsection\label{app:factor_initialization}
The auxiliary panel provides an initial factor estimate: take the leading $r$ right singular vectors of $Z_{\mathcal A}M_1$, where $M_1=I_T-\ind_T\ind_T'/T$, and normalize the factors. At each iteration, let $f_t$ in the following row update denote the current factor value. Each row $j=1,\ldots,qN$ estimates its fixed effect and loadings on its observed dates:
\[
 \begin{pmatrix}\widehat\eta_j\\\widehat\lambda_j\end{pmatrix}
 =\left(\sum_tR_{jt}x_t^fx_t^{f\prime}\right)^{-1}\sum_tR_{jt}x_t^fZ_{jt},
 \qquad x_t^f=(1,f_t')'.
\]
The initial factors give $\widetilde C$. We alternate observed-row regressions and factor updates at each date using all observed rows. Each update transfers the factor mean to the fixed effects and normalizes the centered factors, transforming loadings to preserve fitted values. This separates the fixed effect from the observed factor mean and uses the initialization from sufficient auxiliary information of \citet{DPX2024}.

The common-component estimate is $\widehat C$, and the fitted residuals are $\widehat e=Z-\widehat C$ on observed entries. These estimate the factor-model errors, which differ from the local-projection error.

\item \textbf{Step 2. Restore the full-period projection moments.}
Retain observed values, fill missing entries with $\widehat C$, and restore the missing idiosyncratic covariance contribution. For pair type $k$, put $m=T_h$, $n=NT_h$, and let $p_{ik}$ be its observed fraction for unit $i$. Form
\begin{equation}\label{eq:gammahat}
 \check Z=R\circ Z+(1-R)\circ\widehat C,\qquad
 \widehat\Gamma_{ik}=\frac1{mp_{ik}}\sum_{(u,v)\in\mathcal I_{ik}}R_uR_v\widehat e_u\widehat e_v.
\end{equation}
The moment-restored local projection (MRLP) starts from
\begin{equation}\label{eq:muhat}
 \widehat\mu_k=F_k(\check Z)+\frac1N\sum_i(1-p_{ik})\widehat\Gamma_{ik},
 \qquad \widehat\beta=\widehat Q^{-1}\widehat q.
\end{equation}
Here $F_k$ is the within-unit product over the full admissible period, and $\widehat Q,\widehat q$ assemble the regressor and regressor--outcome moments. An observed pair already contributes its idiosyncratic error product through $F_k(\check Z)$. Only the fraction $1-p_{ik}$ remains to be restored. This construction preserves observed variation and adds the covariance omitted by conditional-mean imputation, a distinction emphasized by \citet{CBN2023}. Covariances are estimated separately for each unit before averaging. Pooling all observed pairs would give greater weight to units with longer histories.

\item \textbf{Step 3. Correct the moments and estimate the coefficients.}
Two biases remain in \eqref{eq:muhat}. Put
$D_R=\operatorname{diag}(\vecop R)$, $\Sigma_e=\operatorname{Var}(\vecop E\mid C,R)$, $S_R=D_R\Sigma_eD_R$ and
$A_k^0=D_R(J_k-W_k)D_R$, where $J_k,W_k$ are defined in
Online Appendix~\ref*{app:score_expansion}. The corrected moments are
\begin{equation}\label{eq:bc}
 \widehat\mu_k^{\mathrm{bc}}=\widehat\mu_k+\widehat a_k-\widehat b_k,
 \qquad
 \boxed{\widehat\beta^{\mathrm{bc}}
   =(\widehat Q^{\mathrm{bc}})^{-1}\widehat q^{\mathrm{bc}}.}
\end{equation}
Here
\begin{equation}\label{eq:mean_correction}
 \widehat a_k=n^{-1}\tr(A_k^0\widehat S_R),\qquad
 b_k=n^{-1}\tr(B_kS_R),\qquad
 \widehat b_k=n^{-1}\tr(\widehat B_k\widehat S_R)+\varepsilon_k^{\mathrm{tr}}.
\end{equation}
The term $\varepsilon_k^{\mathrm{tr}}$ is the centered numerical error of the Gaussian trace average in \begin{NoHyper}\eqref{eq:feasibletrace}\end{NoHyper}. The covariance matrix in these corrections is restricted to observed entries: rows and columns corresponding to missing entries are zero. The first correction restores the mean products subtracted by within-unit
demeaning. The second removes the quadratic effect of fitting the common
component and residual products. The symmetric matrix $B_k$ and its
computation without forming the full matrix are given in
\begin{NoHyper}\eqref{eq:U22}\end{NoHyper}.

For a row pair and lag with $q_{jv,l}\ge k_T$, estimate its covariance
using clipped products of fitted residuals from jointly observed dates divided by $q_{jv,l}$.
Set the estimate to zero for smaller counts and for $|l|>L_T$.
Retain the qualifying within-unit covariances and threshold the
qualifying cross-unit covariances. The resulting covariance estimate on observed entries is $\widehat S_R=D_R\operatorname{Toep}
(\widehat\Gamma^{\mathrm{trim}})D_R$.
Count trimming retains the observed entries and all units in the panel.
Online Appendix~\ref*{app:feasible} gives the formula and functional-consistency
proof. When the observed histories support it, a residual degrees-of-freedom adjustment can be used in the same covariance calculation. Corollary~\ref*{cor:covariance_adjustment} verifies this refinement without a separate covariance assumption in the main results. Independent Gaussian vectors approximate the fitting trace in
\begin{NoHyper}\eqref{eq:feasibletrace}\end{NoHyper}. They do not impose Gaussian idiosyncratic
errors. A~corrected regressor second-moment matrix that is not positive definite
is recorded as an estimation failure.
\end{enumerate}
The projection fixed effects are estimated by $\widehat\alpha_{ih}=\bar{\widehat c}_{y,i}-\bar{\widehat c}_{x,i}'\widehat\beta^{\mathrm{bc}}$, where the bars average fitted conditional means over the projection period. Standard errors are computed separately in Section~\ref{subsec:inference}.

\begin{remark}[The balanced-panel reduction]\label{rem:balanced}
If every entry is observed, $p_{ik}=1$, $\check Z=Z$, and the factor-estimation correction $\widehat b_k$ is identically zero. Equation~\eqref{eq:bc} becomes
\begin{equation}\label{eq:balanced_reduction}
 \widehat\mu_k^{\mathrm{bc}}=F_k(Z)+\widehat a_k.
\end{equation}
The estimator then corrects the demeaning bias in the ordinary within moments. This reduction permits nonzero factor loadings in the outcome and regressors and requires no change of estimator. Factors still help estimate the covariance sequence used in $\widehat a_k$. Retaining observed variation can increase variance relative to replacing all observations by fitted conditional means. The correction does not imply an efficiency ordering. Online Appendix~\ref*{app:balanced} proves the reduction and gives a precise Gaussian variance comparison.
\end{remark}

\begin{remark}[Comparison with MSS]\label{rem:spj_comparison}
Let $\widehat\beta_{h,F}^{\mathrm{FE}}$ be the within estimator using all complete regression cases. Following the observation split in \citet[Appendix~B.2.4]{Mei2026}, split each unit's usable observations chronologically into two halves. Denote the corresponding within estimates by $\widehat\beta_{h,A}^{\mathrm{FE}}$ and $\widehat\beta_{h,B}^{\mathrm{FE}}$. The published benchmark is
\[
 \widehat\beta_h^{\mathrm{MSS}}
 =2\widehat\beta_{h,F}^{\mathrm{FE}}
  -\frac12\bigl(\widehat\beta_{h,A}^{\mathrm{FE}}+
                  \widehat\beta_{h,B}^{\mathrm{FE}}\bigr).
\]
MSS uses observed regression cases without factor estimation or moment restoration. Its unbalanced-panel argument considers consecutive histories with different starting and ending dates. Under their common-coefficient and stationarity conditions, the full and half histories estimate the same slope, and the split removes the leading demeaning bias. Our full-period projection retains the contribution of the realized common factors. When the observed windows give that variation different weights, correcting demeaning bias alone need not recover the target in \eqref{eq:target}. MRLP-BC restores those moments and accounts for their estimation. It can also use pairwise observations when no complete regression case remains. These gains require the factor and covariance restrictions of our model and do not imply that MSS fails merely because a panel is unbalanced.
\end{remark}

\begin{remark}[Relation to componentwise correction]\label{rem:componentwise}
For a scalar projection moment, let $\Sigma=\E[ee'\mid C,R]$ and write its leading centered score as
\[
 Z_m=n^{-1/2}\{\ell_m'e+e'K_me-\tr(K_m\Sigma)\}.
\]
Its conditional variance is exactly
\[
 \operatorname{Var}(Z_m\mid C,R)=n^{-1}
 \{\ell_m'\Sigma\ell_m+2\operatorname{Cov}(\ell_m'e,e'K_me\mid C,R)
                 +\operatorname{Var}(e'K_me\mid C,R)\}.
\]
The middle term need not vanish under asymmetric innovations. The terms also retain cross-unit dependence allowed by Assumption~\ref{ass:dependence}. Equation~(13) of the March 2026 version of \citet{DP2026} likewise corrects moment components using observed covariate pairs. Its panel argument uses cross-fitting across independent units, allowing dependence within each unit. Here the factor fit, within transformation and covariance restoration use the same panel, and the joint expansion carries their contribution through the inverse moment matrix.
\end{remark}

\subsection{Asymptotic distribution}\label{subsec:theory_assumptions}\label{subsec:expansion}
We condition on $C$ and $R$ and center all limits at the finite-period projection $\beta_0=\beta_{h,NT}$. Write $n=NT_h$, where $T_h=T-O(1)$ is the number of admissible projection dates. A limiting coefficient cannot replace this target unless their difference is $o(n^{-1/2})$. For $e_t,\varepsilon_t\in\R^{qN}$, write $e=\vecop(E)$ and $\Sigma_e=\operatorname{Var}(e\mid C,R)$.

\begin{assumption}[Idiosyncratic dependence]\label{ass:dependence}
For fixed $h,q,r,\{\ell_{1k},\ell_{2k}\}_k$, $C<\infty$ and $0<\varrho<1$,
\begin{enumerate}[label=(\roman*)]
\item
\begin{equation}\label{eq:process}
 e_t=\sum_{\ell\ge0}A_{\ell,N}\varepsilon_{t-\ell},\qquad
 \|A_{\ell,N}\|_1+\|A_{\ell,N}\|_\infty\le C\varrho^\ell.
\end{equation}
\item $\{\varepsilon_{ta}:t\in\mathbb Z,\ 1\le a\le qN\}$ mutually independent,
$\E\varepsilon_{ta}=0$, $\E\varepsilon_{ta}^2=1$,
\mbox{$\sup_{N,t,a}\E|\varepsilon_{ta}|^8<\infty$}.
\item $\Sigma_e\preceq CI$, $N/T\to\kappa\in(0,\infty)$.
\end{enumerate}
\end{assumption}
Here $\|\cdot\|_1$ and $\|\cdot\|_\infty$ are maximum absolute column and row sums. Assumption~\ref{ass:dependence}(ii) imposes independence on the innovations, while the filter in Assumption~\ref{ass:dependence}(i) permits heterogeneous variances and serial and cross-unit dependence in the errors. The row and column bounds limit the total strength of dependence and exclude an omitted pervasive shock. Common variation of that kind belongs in the factor component. These bounds permit local cross-unit dependence while restricting its aggregate strength. The covariance sparsity condition below further restricts which weak links can be discarded in feasible inference, following the distinction between dependence and estimability in \citet{BCL2021,BCL2024}.

The geometric time bound and eighth innovation moments control lag truncation and fourth-order residual products. They permit asymmetric innovations, so they do not eliminate the covariance between the linear and quadratic parts of the score. Persistence is fixed below one, and the horizon and factor rank are fixed. Together with $N/T\to\kappa$, these restrictions place the first-stage and demeaning biases on the same $\sqrt{NT}$ scale as the coefficient error. They do not provide the uniform persistence conclusions studied for lag-augmented time-series local projections by \citet{MOPM2021}.

Let $\widetilde C$ be the auxiliary initializer described above and $\widehat P_{\mathcal T}$ the orthogonal projection on the affine rank-$r$ tangent space at $\widehat C$.
\Needspace{7\baselineskip}
\begin{assumption}[Accuracy of the factor fit]\label{ass:fit}\label{app:implementation}
With probability tending to one, the fitted time-varying component has rank $r$, obeys the constraints in \eqref{eq:affinefit}, and satisfies
\begin{equation}\label{eq:fit_accuracy}
 \mathcal J(\widehat C)\le\mathcal J(\widetilde C),\qquad
 \|\widehat P_{\mathcal T}\mathcal P_R(Z-\widehat C)\|_F\le(N+T)^{-1/2}.
\end{equation}
\end{assumption}
The two numerical requirements control different errors. The objective bound compares the joint fit with an initializer whose factor information comes from the complete auxiliaries. The projected-gradient bound controls the remaining error in the fitted normal equations. Lemma~\ref*{lem:P2} establishes the resulting local expansion and existence of a constrained minimizer satisfying these bounds. A computed fit must meet these numerical conditions. The result is not a global convergence claim for an arbitrary alternating-least-squares initialization.

These conditions also matter for inference on the projection. Consistency of imputed values alone does not determine the bias of their products. Factor-augmented regression accounts for estimated factors \citep{BN2006}, and low-rank inference develops expansions for its specified entries or functionals \citep{CHLZ2023,CKL2024,CY2024}. Here the fit must also be expanded to second order inside the restored projection moments. Its first-order error contributes to sampling variation, while its curvature contributes to bias. Both terms are derived for the observed-entry fit used by the estimator, without an independent training sample.

Feasible bias correction estimates covariance functionals on the observed entries. Let $L_T$ be the nuisance-covariance lag cutoff and $\tau_\Sigma$ its cross-unit threshold. For the clipping and threshold scales $\widehat s_{jv}$, require
\[
 \inf_{j,v}\widehat s_{jv}\ge s_0>0,\qquad
 \sup_{j,v}\widehat s_{jv}=O_p(1).
\]
For $\varrho<\varrho_0<1$, define
\begin{equation}\label{eq:summability}
 D_N=\sum_{\ell\ge0}\varrho_0^{-\ell}|A_{\ell,N}|,\qquad
 \vartheta_{ij}=\sum_{a,b=1}^q(D_ND_N')_{(i,a),(j,b)},\qquad
 H_N(u)=\max_i\sum_{j\ne i}\min(\vartheta_{ij},u).
\end{equation}
The absolute value is entrywise. The next restriction controls the sum of small covariances discarded by thresholding.
For nuisance covariances, write
$q_{jv,l}=\sum_{t-s=l}R_{jt}R_{vs}$, let $k_T$ be a count cutoff,
and define
\[
a_T^*=T^{1/4}k_T^{-1/2}+\log^2(NT)k_T^{-1/2}+T^{-1/2}.
\]
\begin{assumption}[Covariance estimation]\label{ass:feasible}
\begin{enumerate}[label=(\roman*)]
\item $\lim_{u\downarrow0}\limsup_NH_N(u)=0$.
\item $L_T\to\infty$, $L_T=O\{\log(NT)\}$.
\item $k_T\to\infty$, $k_T/T\to0$, $\tau_\Sigma\to0$, and
\[
 a_T^*/\tau_\Sigma+L_T\sqrt{\tau_\Sigma}\to0.
\]
\end{enumerate}
\end{assumption}

Assumption~\ref{ass:feasible}(i) controls the aggregate contribution of small cross-unit covariances discarded by thresholding, as in \citet{BCL2021,BCL2024}. Assumption~\ref{ass:feasible}(ii) lets the lag cutoff increase while limiting the estimation error accumulated across lags. Assumption~\ref{ass:feasible}(iii) permits nuisance covariance pairs with few or no common observations. The count cutoff $k_T$ omits their estimates, and the weights on observed entries make their aggregate contribution vanish when $k_T/T\to0$. For example, $k_T=\lceil c_kT^{3/4}\rceil$, $\tau_\Sigma=c_\Sigma T^{-1/16}$ and $L_T=\lceil c_L\log T\rceil$ satisfy the rates for fixed positive constants. Lemma~\ref*{lem:rare} bounds the omitted contribution.

Count trimming removes estimates supported by too few observations, while thresholding removes small cross-unit covariances that would otherwise accumulate estimation noise. Both operations retain the units and observed outcomes. When every nuisance pair has a positive limiting overlap fraction, count trimming is eventually inactive. The identifying pairs in Assumption~\ref{ass:pairs} remain a separate requirement.

Lag truncation, covariance thresholding and clipping have precedents in \citet{NW1987,NW1994,BCL2021,BCL2024,FWZ2021}. The rates established for fitted residuals control the covariance functionals used in the correction. The covariance of the leading score is estimated separately by the localized multiplier bootstrap in Section~\ref{subsec:inference}. Online Appendix Remark~\ref*{app:regularization_discussion} discusses the regularization choices and alternatives.

The leading score combines linear terms from observed values and estimated conditional means with quadratic terms from observed residual products. To define it, let $G_k$ satisfy $\mathrm DF_k(C)[X]=\langle G_k,X\rangle/n$, let $P$ project onto the affine factor tangent space, and let $H_0=(P\mathcal P_RP)|_{\mathcal T}$. Put
\[
 d_k=H_0^{-1}P\{(1-R)\circ G_k\},\qquad
 \ell_k=\vecop\{R\circ(G_k+d_k)\}.
\]
Let $K_k$ be the symmetric matrix satisfying $e'K_ke=\sum_{i,(u,v)\in\mathcal I_{ik}}(R_uR_v/p_{ik})e_ue_v$. If $c_{jk}(\beta_0)$ assembles moment $k$ into normal equation $j$, set $\ell_j=\sum_kc_{jk}(\beta_0)\ell_k$ and $K_j=\sum_kc_{jk}(\beta_0)K_k$. Online Appendix~\ref*{app:score_expansion} gives the explicit tangent projection and matrix formulas.

\begin{theorem}[Coefficient expansion and asymptotic normality]\label{thm:expansion}
Let $n=NT_h$, $\beta_0=Q_{NT}^{-1}q_{NT}$, and
$\widehat\beta^{\mathrm{bc}}=(\widehat Q^{\mathrm{bc}})^{-1}\widehat q^{\mathrm{bc}}$
be defined by \eqref{eq:bc}. For $e=\vecop(Z-C)$ and the weights $\ell_j,K_j$ defined above, set
\[
 \psi_{NT,j}=n^{-1/2}\{\ell_j'e+e'K_je-\tr(K_j\Sigma_e)\},\qquad
 \Sigma_e=\operatorname{Var}(e\mid C,R),\quad
 \Omega_{NT}=\operatorname{Var}(\psi_{NT}\mid C,R).
\]
Under Assumptions~\ref{ass:calendar}--\ref{ass:feasible},
\[
 \|\widehat Q^{\mathrm{bc}}-Q_{NT}\|=o_p(1),\qquad
 \Pr\{\lambda_{\min}(\widehat Q^{\mathrm{bc}})>0\mid C,R\}\longrightarrow1,
\]
\begin{equation}\label{eq:mainexpansion}
 \sqrt n(\widehat\beta^{\mathrm{bc}}-\beta_0)=Q_{NT}^{-1}\psi_{NT}+o_p(1).
\end{equation}
If $\Omega_{NT}\to\Omega\succ0$, then, with $Q=\lim Q_{NT}\succ0$,
\[
 \sqrt{NT_h}(\widehat\beta^{\mathrm{bc}}-\beta_{h,NT})
 \Rightarrow N(0,Q^{-1}\Omega Q^{-\prime}).
\]
\end{theorem}
The expansion separates the effect of missing observations on centering from their effect on uncertainty. The corrections remove biases of order $N^{-1}+T^{-1}$ without adding first-order sampling variation. When $N/T$ has a positive finite limit, leaving these biases uncorrected shifts the limiting distribution at the standard-error scale. This is the panel bias problem underlying analytical and jackknife corrections \citep{HN2004,FVW2016,Mei2026}, with additional terms here because the missing contributions are themselves estimated.

With complete observations, $d_k=0$ and $\ell_k=\vecop G_k$. The second-order factor-estimation correction then disappears exactly, while the demeaning correction remains. With missing entries, $d_k$ carries factor-estimation uncertainty into the leading score. Thus correction of first-stage bias does not make the estimated missing values equivalent to known values for inference. Relative to imputation-based inference methods using cross-fitting \citep{DP2026}, the theorem accounts directly for the joint estimation on the same dependent panel.

Inference uses the covariance of the full score in the expansion. The linear and centered quadratic terms can be correlated through third innovation moments. Gaussian quadratic-form formulas would omit this interaction under asymmetric innovations. DGP~3 preserves the relevant second moments while changing asymmetry, so that this feature can be assessed without changing the population projection. Online Appendix~\ref*{app:main_proof} derives the joint central limit theorem from the innovation conditions.

\subsection{Conditional projections with fully observed regressors}\label{subsec:conditional}
Suppose the regressor paths $X$ are fully observed and the outcome is partly observed. We consider the projection of the conditional outcome mean on $X$, conditioning throughout on $(C,R,X)$. The outcome and auxiliary variables are fitted jointly. Their panel has $J\asymp N$ rows, with at most $q_*<\infty$ rows per unit, and the auxiliary units may differ from the $N$ target units. Assumptions~\ref{ass:calendar}, \ref{ass:dependence}, \ref{ass:fit} and~\ref{ass:feasible} apply with $J$ replacing $qN$ and with probability conditional on $(C,R,X)$. The complete set $\mathcal A$ need not have $q_0N$ rows but retains the stated auxiliary loading Gram bound. The corresponding proofs use only $J\asymp N$ and the bounded number of rows per unit.

Write $x_{it}^{w,h}=x_{it}-T_h^{-1}\sum_{s\in\mathcal T_h}x_{is}$ and
\[
 Q_h^X=n^{-1}\sum_{i,t\in\mathcal T_h}x_{it}^{w,h}x_{it}^{w,h\prime},
 \qquad \beta_h^C=(Q_h^X)^{-1}n^{-1}
          \sum_{i,t\in\mathcal T_h}x_{it}^{w,h}C_{i,t+h}.
\]
\begin{assumption}[Fully observed regressors]\label{ass:conditional}
\[
 \sup_{i,t}\|x_{it}\|\le C,\qquad
 \lambda_{\min}(Q_h^X)\ge c>0,\qquad
 \E[E\mid C,R,X]=0.
\]
\end{assumption}

Full observation of $X$ and conditional mean independence are separate requirements. Assumption~\ref{ass:conditional} conditions on the entire regressor path, including values after a given projection date. Orthogonality to current and past regressors alone does not imply this restriction. Under the stated conditioning, $Q_h^X$ is fixed and the idiosyncratic outcome contribution has mean zero. The target $\beta_h^C$ is consequently a projection of the conditional mean over the specified period.

For a fixed contrast $\theta_h=v'\beta_h^C$ with $v\ne0$, $\widehat\theta_h$ projects the completed outcome on the observed regressors. The contrast weights $\ell_h$ and fitting-bias matrix $B_h$ are defined in Online Appendix~\ref*{app:conditional_weights}.

Under Assumption~\ref{ass:conditional}, the regressor Gram matrix is observed and the idiosyncratic outcome error has conditional mean zero. Corrections for missing idiosyncratic regressor--outcome covariances and for demeaning are therefore unnecessary. Estimating the missing outcome means still introduces second-order bias and first-stage uncertainty.
Define
\[
 \widehat\theta_h^{\mathrm{bc}}
 =\widehat\theta_h-n^{-1}\tr(\widehat B_h\widehat S_R),\qquad
 \widehat s_h^2=n^{-2}\widehat\ell_h'\widehat S_R\widehat\ell_h.
\]
\begin{theorem}[Conditional MRLP-BC]\label{thm:conditional}
Under Assumptions~\ref{ass:calendar}, \ref{ass:dependence},
\ref{ass:fit}, \ref{ass:feasible} and~\ref{ass:conditional}, if
$\ell_h'S_R\ell_h/n\to\omega_h>0$,
\begin{enumerate}[label=(\roman*)]
\item
\[
 \sqrt n(\widehat\theta_h^{\mathrm{bc}}-\theta_h)
 =n^{-1/2}\ell_h'e+o_p(1).
\]
\item
\[
 (\widehat\theta_h^{\mathrm{bc}}-\theta_h)/\widehat s_h
 \Rightarrow N(0,1).
\]
\end{enumerate}
\end{theorem}

Theorem~\ref{thm:conditional} has a linear leading score because the regressor Gram matrix is fixed under the stated conditioning.\footnote{The proof and contrast weights are in Online Appendix~\ref*{app:conditional_proof}.} Estimating the missing conditional means still affects the leading weights $\ell_h$ and introduces the second-order term $B_h$. A fixed positive number of independent Gaussian draws can evaluate the correction trace without changing the first-order limit. The variance plug-in uses $\widehat S_R$. Section~\ref{subsec:inference} instead estimates the covariance of the complete score by the localized multiplier bootstrap.

The same construction allows a prespecified set of dates for each unit. If $W_{it}$ indicates whether an outcome date is included, use
\[
 A_{it,h}=W_{it}W_{i,t+h}\ind(1\le t\le T-h),\qquad
 n_h^W=\sum_{it}A_{it,h}.
\]
Within-unit means and projection weights are computed on these dates. The resulting coefficient describes the projection over those dates, with normalization $n_h^W$ in place of $NT_h$. Proposition~\ref*{prop:known_windows} and Corollary~\ref*{cor:window_lmb} give its expansion and bootstrap inference under the stated restrictions conditional on the selected dates.

\begin{remark}
The conditional restrictions apply to the population specified in the
projection. Selecting units using noisy auxiliary observations can add
information beyond $(C,R,X)$. Conditional mean zero and the innovation
restrictions need not be preserved after conditioning on that selection.
Treating the realized membership and rank as fixed does not establish
coverage conditional on selection or under repeated selection.
\end{remark}

\subsection{Localized multiplier bootstrap}\label{subsec:inference}

The localized multiplier bootstrap (LMB) applies correlated Gaussian weights to the score in the coefficient expansion. Allocate each fitted linear term to its cross-sectional unit and actual date, and each restored residual product to its unit and later date. Center a residual product by its mean over the same observed pairs. Combining these terms at $\widehat\beta^{\mathrm{bc}}$ gives $\widehat\xi_{it}$, the estimated counterpart of the score in \begin{NoHyper}\eqref{eq:I12}\end{NoHyper}. Let $\mathcal J_i$ collect all fitted-panel rows belonging to cross-sectional unit $i$. Target and auxiliary rows of the same unit share its multiplier weight. The number of represented units is $N_U\asymp N$, and the projection normalization remains $n=NT_h$.

Let $a_{it}\in\mathbb R^{d_N}$ describe available dependence information. The vectors may vary over time. Enlarge the conditioning information of the applicable coefficient expansion by their histories, maintaining the innovation and factor-estimation restrictions under this conditioning. Prior dating alone does not establish these restrictions. With $\delta=\min(\sqrt N,\sqrt T)$ and $M=N_UT\asymp n$, define
\begin{align}
 \nu_{NT}&=\max_{t,s}\max_i\sum_j|a_{it}'a_{js}|,\label{eq:dated_nu}\\
 \mathcal E_{NT}&=\frac1M\sum_{i,t,j,s}r_{it}r_{js}
 |1-a_{it}'a_{js}|\vartheta^U_{ij}
 \varrho_0^{(|t-s|-2H)_+}.\label{eq:dated_loss}
\end{align}
Here $r_{it}=\ind\{\sum_{a\in\mathcal J_i}R_{ita}>0\}$ indicates that at least one row of unit $i$ is observed at date $t$, $\vartheta^U$ is the innovation-dependence envelope grouped by cross-sectional units, and $H$ bounds the fixed time shifts in the score. Their construction is given in Online Appendix~\ref*{app:lmb}.

\begin{assumption}[Localization]\label{ass:dated_localization}
\begin{enumerate}[label=(\roman*)]
\item $a_{it}$ is fixed under the stated conditioning and $\|a_{it}\|=1$.
\item $\ell_T\in\mathbb N$, $\ell_T\to\infty$, $\nu_{NT}\ell_T/\delta\to0$.
\item $\mathcal E_{NT}\to0$.
\end{enumerate}
\end{assumption}

Assumption~\ref{ass:dated_localization}(iii) requires the weights to retain the material score covariances on average over the observed support. Positive semidefiniteness alone does not imply this approximation. Assumption~\ref{ass:dated_localization}(ii) limits the accumulation of estimation error in the fitted score. A fixed number of groups whose sizes grow proportionally to $N$ does not satisfy this sufficient condition.

Conditionally on the data, draw a Gaussian field satisfying
\begin{equation}\label{eq:dated_weights}
 \E^*\omega_{it}^*=0,\qquad
 \E^*(\omega_{it}^*\omega_{js}^*)
   =(a_{it}'a_{js})\left(1-\frac{|t-s|}{\ell_T}\right)_+.
\end{equation}
The covariance is positive semidefinite because it is a Gram matrix. Apply the same field across the reported horizons. The two constructions are
\begin{enumerate}[label=(\roman*)]
\item For the general projection,
\[
 G^*=n^{-1/2}\sum_{i,t}\widehat\xi_{it}\omega_{it}^*,\qquad
 \widehat\beta^*=\widehat\beta^{\mathrm{bc}}
       +n^{-1/2}(\widehat Q^{\mathrm{bc}})^{-1}G^*.
\]
\item For the conditional contrast with fully observed regressors,
\[
 G_h^*=n^{-1/2}\sum_{i,t}\widehat\xi_{it,h}\omega_{it}^*,\qquad
 \widehat\theta_h^*=\widehat\theta_h^{\mathrm{bc}}+n^{-1/2}G_h^*,
 \widehat\xi_{it,h}=\sum_{a\in\mathcal J_i}\widehat\ell_{h,ita}\widehat e_{ita}.
\]
\end{enumerate}
In the conditional construction, $\Omega_{NT}=\operatorname{Var}(n^{-1/2}\ell_h^{\prime}e\mid C,R,X)$ and the scalar inverse moment factor is one. The inverse regressor Gram matrix is already included in $\widehat\ell_h$ in the second construction. The factor fit, observation pattern and analytical correction are held fixed in the multiplier draws. Their estimation effects enter through the score. Related kernel weights appear in the spatial dependent wild bootstrap of \citet{CGKP2023} and the dependent wild bootstrap of \citet{Shao2010DWB}. The result below controls substitution of the present fitted score.

Let
\[
 \widehat V^{\mathrm{LMB}}=(\widehat Q^{\mathrm{bc}})^{-1}
 \E^*(G^*G^{*\prime})(\widehat Q^{\mathrm{bc}})^{-\prime}.
\]
\begin{theorem}[Localized multiplier bootstrap]\label{prop:dated_covariance}
Under the applicable conditions for the coefficient expansion,
Assumption~\ref{ass:dated_localization} and the conditions on innovations and estimated scores
in Lemmas~\ref*{lem:lmb_dependence} and~\ref*{lem:lmb_score_error},
\begin{enumerate}[label=(\roman*)]
\item
\[
\begin{split}
 \|\E^*(G^*G^{*\prime})-\Omega_{NT}\|
 =O_p\bigg\{\mathcal E_{NT}+\ell_T^{-1}
 +\sqrt{\frac{\nu_{NT}\ell_T}{n}}\hspace{20pt}\\
 +\sqrt{\frac{\nu_{NT}\ell_T}{\delta}}
 +\frac{\nu_{NT}\ell_T}{\delta}\bigg\}=o_p(1).
\end{split}
\]
\item If $Q_{NT}\to Q\succ0$, $\Omega_{NT}\to\Omega\succeq0$ and
$v'Q^{-1}\Omega Q^{-\prime}v>0$,
\[
 \frac{\sqrt n\,v'(\widehat\beta^{\mathrm{bc}}-\beta_0)}
      {(v'\widehat V^{\mathrm{LMB}}v)^{1/2}}
 \Rightarrow N(0,1).
\]
\end{enumerate}
\end{theorem}

The estimated coefficient covariance is $n^{-1}\widehat V^{\mathrm{LMB}}$, so a marginal interval uses standard error $\{\widehat V^{\mathrm{LMB}}_{jj}/n\}^{1/2}$. Theorem~\ref{prop:dated_covariance} retains the covariance restrictions needed for bias correction and replaces thresholding and clipping of score products. The proof bounds bias and sampling variation for the covariance estimator based on population scores, then controls the error from estimating those scores. It is given in Online Appendix~\ref*{app:dated_localization}.

For the conditional contrast, the inverse regressor Gram matrix is already included in $\ell_h$, so the result applies with $Q=1$. Shared draws give simultaneous bands over a fixed finite set of horizons with positive marginal limiting variances. Proposition~\ref*{prop:lmb_bands} gives their ordered-multiplier critical value. For a single contrast, the Gaussian bootstrap interval equals the interval based on its kernel covariance. The bands concern the stated population, without testing differences between populations. The thresholded-score alternative is in Online Appendix~\ref*{subsec:thresholded_inference}.

\begin{remark}[Two-way clustered standard errors]\label{rem:primary_clustering}
Two-way clustering applies to valid regression scores under the conditions of \citet{Thompson2011}. Residuals from a completed regression alone omit the common-component derivative and the residual products restoring missing moments. Fitted row normal equations can also set unit score totals to zero. Lemma~\ref*{lem:lmb_counterexample} shows how clustering such fitted totals can lose variation present in the estimator. This concerns substitution of an estimated score, rather than a failure of ordinary-regression clustering. Thompson's lagged extension permits cross-unit covariance across dates under its conditions. The localized multiplier bootstrap controls substitution of estimated scores through Assumption~\ref{ass:dated_localization}.
\end{remark}

\section{Monte Carlo Simulations}\label{sec:simulations}
The simulations examine recovery of the intended projection, correction of demeaning bias and the coverage of bootstrap confidence intervals. These questions require different comparisons. DGPs~1--4 allow the observed histories to identify a different projection from the full period. DGP~5 holds the coefficient constant across histories to compare bias corrections directly. DGP~6 studies conditional projections with fully observed regressors and quarterly outcomes on a monthly grid. Table~\ref{tab:simulation_dgps} summarizes the designs.

\begin{table}[htp!]\centering\small
\caption{Simulation models and observation patterns}\label{tab:simulation_dgps}
\setlength{\tabcolsep}{4pt}\renewcommand{\arraystretch}{1.08}
\begin{tabular*}{\linewidth}{@{\extracolsep{\fill}}lp{0.34\linewidth}>{\raggedright\arraybackslash}p{0.24\linewidth}>{\raggedright\arraybackslash}p{0.25\linewidth}@{}}\toprule
Design & Model component & Factors and innovations & Observation patterns\\\midrule
DGP 1 & $v_{it}=\sigma_i(\varepsilon^d_{it}+0.5\varepsilon^d_{i,t-1})/\sqrt{1.25}$ & One factor, Gaussian & Periodic gaps, blocks\\[4pt]
DGP 2 & $v_{it}=0.5v_{i,t-1}+\sqrt{0.75}\,\sigma_i\varepsilon^d_{it}$ & One factor, Gaussian & Periodic gaps, blocks, random, internal gaps, selection on the common component, balanced\\[4pt]
DGP 3 & As in DGP 1 & One factor, $(\chi^2_3-3)/\sqrt6$ & Periodic gaps, blocks, random, internal gaps, selection on the common component\\[4pt]
DGP 4 & As in DGP 2 & Two factors, Gaussian & Periodic gaps, blocks\\[4pt]
DGP 5 & $v_{it}=\rho v_{i,t-1}+\sqrt{1-\rho^2}\,\sigma_i\varepsilon^d_{it}$ & One auxiliary factor, Gaussian, $\rho\in\{0.5,0.8\}$ & Balanced, unequal consecutive histories\\[4pt]
DGP 6 & $x_{it}=(D_i f_{1t},f_{1t})'$ fully observed & Three factors, centered exponential errors & Random, blocks, internal gaps, mixed frequency\\\bottomrule
\end{tabular*}\par\smallskip
\begin{minipage}{\linewidth}\tablenotefont
Notes: DGPs 1--4 retain factor variation in the projection. DGP 5 has zero target factor loadings and a common coefficient in every observation window. DGP 6 uses the conditional projection with fully observed regressors. Baseline innovations are independent across units in DGPs 1--5. DGPs 2 and 3 also have a nearest-neighbor dependence extension. DGP 6 includes serial and cross-unit dependence throughout. All designs have complete informative auxiliaries and informative observed factor variation. The text defines the observation patterns, dimensions and horizons. Factor-based methods receive the true rank, while MSS does not use factors.
\end{minipage}\end{table}

\subsection{Simulation design}\label{subsec:simulation_dgp}
In DGPs~1--4, we project $y_{i,t+h}$ on $(d_{it},d_{i,t-1})'$ and individual fixed effects at $h\in\{1,4,7,10\}$. The underlying series satisfy
\begin{align*}
 y_{it}&=\eta_{iy}+\lambda_{iy}'f_t+v_{i,t-1}+\sigma_i\varepsilon^y_{it},\\
 d_{it}&=\eta_{id}+\lambda_{id}'f_t+v_{it},\\
 a_{it}&=\eta_{ia}+\lambda_{ia}'f_t+\sigma_i\varepsilon^a_{it}.
\end{align*}
The auxiliary series $a_{it}$ is complete. DGP~1 has a moving-average predictor error, and DGP~2 replaces it with an autoregression. DGP~3 introduces asymmetric innovations while preserving the second moments, isolating the higher-moment terms in the score covariance. DGP~4 adds a second factor while preserving each series' common-component variance. Factor variation remains part of the projection in every design.

Periodic gaps remove one date out of every three. No regression case is complete at the reported horizons, although the pairs required for the normal equations remain observed. The block design instead gives two groups consecutive histories covering the first or last two-thirds of the period. Additional scenarios introduce random missingness, internal gaps and selection on the common component. We evaluate each estimator against the same full-period projection, calculated from the common components and idiosyncratic covariances.

DGP~5 adapts the predictive system of \citet{Mei2026} and removes target factor loadings. With $\theta=-0.6$,
\[
 y_{it}=\eta_{iy}+\theta v_{it}+\sigma_i\varepsilon^y_{it},\qquad
 d_{it}=\eta_{id}+v_{it},\qquad
 v_{it}=\rho v_{i,t-1}+\sqrt{1-\rho^2}\,\sigma_i\varepsilon^d_{it}.
\]
The coefficient is $(\theta\rho^h,0)'$ for $\rho\in\{0.5,0.8\}$ in every full, observed and split history. The auxiliary series retains one factor. Unequal consecutive histories therefore permit a comparison with MSS on a shared coefficient. DGP~6 uses three factors and fully observed regressors $x_{it}=(D_if_{1t},f_{1t})'$, where $D_i$ indicates exposure. Its target is the coefficient of the exposure interaction in the full-period conditional projection. The four observation patterns have the same count within each firm, so differences in performance reflect when observations are available.

DGPs~1 and 2 use $N,T\in\{120,240\}$, and DGPs~3 and 4 use the two square panels. The main comparisons for DGPs~5 and 6 use $N\in\{200,1000\}$ and $T\in\{120,360\}$. In DGP~6, dates are months, with 360 months as the primary calendar. All designs retain strong auxiliary factor information. Common components and observation patterns are held fixed across disturbance replications. The complete distributions, observation schedules and covariance choices are in Online Appendix~\ref*{app:simulation_design}.

\phantomsection\label{subsec:simulation_methods}
MRLP and MRLP-BC use the same joint factor fit and retain the same observed values, so their comparison isolates the corrections. The auxiliary principal components analysis (PCA) benchmark estimates factors from complete auxiliaries and uses $F_k(\widehat C)+N^{-1}\sum_i\widehat\Gamma_{ik}$ in DGPs~1--4. This combines sufficient auxiliary information \citep{DPX2024} with covariance restoration \citep{CBN2023}, without optimal Target-PCA weighting. In DGP~6, auxiliary PCA retains observed outcomes and fills missing ones using affine row regressions. FE uses complete regression cases. MSS uses the split-panel correction in Remark~\ref{rem:spj_comparison}. Full-data FE observes the panel before masking and is an infeasible information benchmark. Factor-based methods receive the true rank, but estimate the factors and loadings.

We report bias, RMSE and the coverage and length of pointwise 95\% intervals. Pooled RMSE averages squared errors across both coefficients and replications before taking the square root. DGP~6 instead reports RMSE for its exposure-interaction coefficient. Methods are evaluated on the same simulated samples, and MRLP-BC intervals include numerical trace uncertainty.\footnote{Figure~\ref{fig:unbalanced_rmse} and Table~\ref{tab:unbalanced_predictive} use 1,000 replications. Table~\ref{tab:conditional_cases} uses 500 matched replications. The Monte Carlo standard error at 95\% coverage is 0.69 percentage points with 1,000 replications and 0.97 with 500. Additional coverage comparisons state their denominators in Online Appendix Table~\ref*{tab:localized_patterns}.}

\subsection{Moment recovery and bias correction}\label{subsec:horizon_results}
\begin{figure}[htp!]\centering
\caption{Pooled RMSE across horizons in unbalanced panels}\label{fig:unbalanced_rmse}
\includegraphics[width=\linewidth]{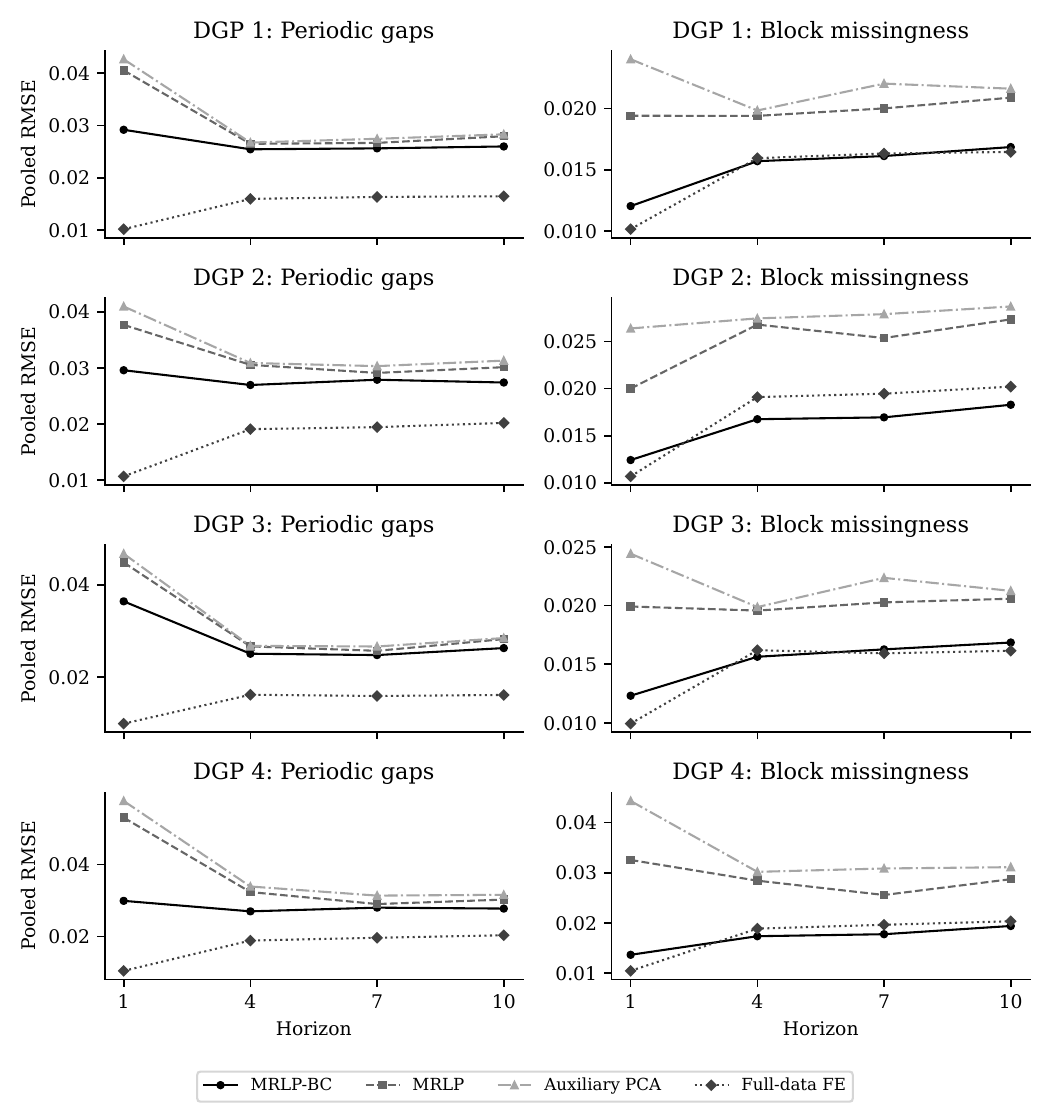}\par\smallskip

\begin{minipage}{\linewidth}\figurenotefont
Notes: $N=T=120$, 1,000 replications, the same full-period target and simulated observations for all feasible methods. Columns distinguish the observation patterns. Rows distinguish DGPs. The specified periodic pattern leaves no complete regression cases. Full-data FE observes the entries withheld from the other procedures and is included only as an information benchmark.\end{minipage}\end{figure}

Moment restoration improves estimation of the full-period projection in the periodic and block designs. MRLP-BC has lower pooled RMSE than MRLP and auxiliary PCA in the reported comparisons, including the periodic scenarios with no complete regression case. Figure~\ref{fig:unbalanced_rmse} illustrates these gains at $N=T=120$. In the two-factor block design at horizon ten, pooled RMSE is 0.0178 for MRLP-BC and 0.0268 for MRLP. The current-predictor coefficient is 0.0007 over the full period but $-0.0156$ over the observed windows. The difference illustrates why correcting observed-case demeaning bias alone need not recover the intended projection.

When the histories identify a common coefficient, MSS and MRLP-BC have similar accuracy with longer panels. In Table~\ref{tab:unbalanced_predictive}, at $N=200$, $T=360$, $\rho=0.5$ and $h=4$, both have pooled RMSE 0.0065. Current-coefficient coverage is 95.7\% for MSS and 95.5\% for MRLP-BC. The corrections matter as the number of firms grows: at $N=1000$ with the same $T$, $\rho$ and $h$, uncorrected FE has bias 0.0043 and coverage 69.5\%. Both corrections reduce bias to 0.0001, with coverage 94.6\% for MSS and 93.7\% for MRLP-BC.

The shorter panel reveals a tradeoff between estimation error and coverage. At $N=1000$, $T=120$, $\rho=0.8$ and $h=10$, MRLP-BC has pooled RMSE 0.0082 against 0.0100 for MSS. Its lagged-coefficient coverage is 93.0\%, compared with 83.4\% for MSS. Current-coefficient coverage is lower, at 92.9\% against 94.4\%. Thus smaller RMSE does not imply uniformly better inference. The balanced version of this comparison is discussed in Online Appendix~\ref*{app:balanced_simulation}.

\begin{table}[htp!]\centering\small
\caption{Unbalanced predictive panels with a common coefficient}\label{tab:unbalanced_predictive}
\setlength{\tabcolsep}{2pt}
\begin{tabular*}{\linewidth}{@{\extracolsep{\fill}}rrrrrrrrr@{}}\toprule
 & & & \multicolumn{2}{c}{Pooled RMSE} & \multicolumn{2}{c}{$d_t$ coverage} & \multicolumn{2}{c}{$d_{t-1}$ coverage}\\
\cmidrule(lr){4-5}\cmidrule(lr){6-7}\cmidrule(l){8-9}
$N$ & $\rho$ & $h$ & MSS & MRLP-BC & MSS & MRLP-BC & MSS & MRLP-BC\\\midrule
\multicolumn{9}{l}{\emph{$T=360$}}\\[2pt]
200 & 0.5 & 4 & 0.0065 & 0.0065 & 95.7 & 95.5 & 95.2 & 95.8\\
200 & 0.5 & 10 & 0.0067 & 0.0067 & 94.6 & 93.9 & 94.3 & 94.0\\
200 & 0.8 & 4 & 0.0092 & 0.0092 & 95.3 & 96.0 & 95.0 & 95.9\\
200 & 0.8 & 10 & 0.0097 & 0.0096 & 95.3 & 94.8 & 96.3 & 95.6\\
\addlinespace
1,000 & 0.5 & 4 & 0.0030 & 0.0031 & 94.6 & 93.7 & 94.3 & 93.9\\
1,000 & 0.5 & 10 & 0.0030 & 0.0030 & 94.7 & 94.1 & 95.3 & 94.7\\
1,000 & 0.8 & 4 & 0.0043 & 0.0043 & 95.2 & 94.9 & 95.2 & 95.2\\
1,000 & 0.8 & 10 & 0.0045 & 0.0044 & 93.7 & 93.4 & 94.3 & 94.1\\
\addlinespace
\multicolumn{9}{l}{\emph{$T=120$}}\\[2pt]
200 & 0.5 & 4 & 0.0117 & 0.0118 & 95.3 & 93.6 & 95.3 & 93.4\\
200 & 0.5 & 10 & 0.0123 & 0.0120 & 95.7 & 94.7 & 94.3 & 93.6\\
200 & 0.8 & 4 & 0.0170 & 0.0169 & 94.2 & 93.3 & 95.6 & 94.8\\
200 & 0.8 & 10 & 0.0184 & 0.0171 & 95.9 & 94.4 & 93.0 & 95.0\\
\addlinespace
1,000 & 0.5 & 4 & 0.0055 & 0.0054 & 95.3 & 93.9 & 93.9 & 93.4\\
1,000 & 0.5 & 10 & 0.0062 & 0.0055 & 93.7 & 93.6 & 89.2 & 94.8\\
1,000 & 0.8 & 4 & 0.0081 & 0.0077 & 93.4 & 94.1 & 94.1 & 93.9\\
1,000 & 0.8 & 10 & 0.0100 & 0.0082 & 94.4 & 92.9 & 83.4 & 93.0\\
\bottomrule\end{tabular*}\par\smallskip
\begin{minipage}{\linewidth}\tablenotefont
Notes: DGP~5, 1,000 Monte Carlo replications per $(N,T,\rho)$ setting. $N$ counts firms, each with one complete auxiliary series. Both methods use the same observed entries and are evaluated against $(-0.6\rho^h,0)'$. MSS uses complete regression cases in each firm's consecutive history. MRLP-BC also uses the auxiliary series and estimates the common component with rank one. Pooled RMSE averages squared errors across both coefficients before taking the square root. Coverage is in percent for pointwise 95\% intervals. MSS uses individual-cluster standard errors and MRLP-BC uses the localized multiplier bootstrap score with independent-unit weights and a logarithmic temporal window. Numerical trace variance is retained. All reported intervals were available. The Monte Carlo standard error at 95\% coverage is 0.69 percentage points.
\end{minipage}\end{table}

Recovering missing outcomes also improves estimation when regressors are fully observed. In DGP~6 at $T=360$ and $h=6$, MRLP-BC has RMSE 0.0036 with 200 firms, compared with 0.0113 for complete-case FE. With 1,000 firms, the values are 0.0016 and 0.0051, while auxiliary PCA has RMSE 0.0017. MRLP and MRLP-BC have similar RMSE, so most of the gain in these cases comes from recovering missing outcomes. The advantage is not universal. At impact with 200 firms, FE has RMSE 0.0104 against 0.0107 for MRLP-BC. A common-coefficient control in Online Appendix~\ref*{app:simulation_design} further examines recovery when observed histories already identify the target.

\begin{table}[htp!]\centering\small
\caption{Conditional monthly projections with mixed-frequency observations}\label{tab:conditional_cases}
\setlength{\tabcolsep}{2pt}
\begin{tabular*}{\linewidth}{@{\extracolsep{\fill}}rrrrrrrrr@{}}\toprule
 & & \multicolumn{4}{c}{RMSE} & \multicolumn{3}{c}{MRLP-BC intervals}\\
$N$ & $h$ & FE & \shortstack{Auxiliary\\PCA} & MRLP & MRLP-BC & \shortstack{Plug-in\\coverage} & \shortstack{LMB\\coverage} & \shortstack{LMB\\mean length}\\\midrule
200 & 0 & 0.0104 & 0.0118 & 0.0108 & 0.0107 & 99.2 & 96.0 & 0.0431\\
200 & 6 & 0.0113 & 0.0037 & 0.0036 & 0.0036 & 99.2 & 95.4 & 0.0146\\
200 & 12 & 0.0102 & 0.0041 & 0.0039 & 0.0039 & 98.6 & 96.0 & 0.0157\\
200 & 18 & 0.0106 & 0.0040 & 0.0038 & 0.0038 & 97.2 & 94.8 & 0.0144\\
1,000 & 0 & 0.0054 & 0.0047 & 0.0046 & 0.0046 & 99.8 & 95.6 & 0.0191\\
1,000 & 6 & 0.0051 & 0.0017 & 0.0016 & 0.0016 & 97.8 & 96.2 & 0.0065\\
1,000 & 12 & 0.0051 & 0.0019 & 0.0017 & 0.0017 & 99.0 & 95.8 & 0.0070\\
1,000 & 18 & 0.0077 & 0.0018 & 0.0016 & 0.0016 & 99.4 & 95.6 & 0.0064\\
\bottomrule\end{tabular*}\par\smallskip
\begin{minipage}{\linewidth}\tablenotefont Notes: DGP~6, $T=360$ months and 500 matched Monte Carlo replications per sample size. RMSE concerns the coefficient of the exposure interaction, evaluated against the same full-period conditional projection in every method. Coverage is in percent for nominal 95\% pointwise MRLP-BC intervals. Both covariance calculations retain the same point estimates and analytical corrections. LMB denotes the localized multiplier bootstrap, with the covariance construction described in Online Appendix~\ref*{app:simulation_design}. Marginal numerical trace variance is included. Every estimate and interval was available. The Monte Carlo standard error at 95\% coverage is 1.0 percentage point.
\end{minipage}\end{table}

\subsection{Bootstrap confidence intervals}\label{subsec:general_simulation}
The localized multiplier bootstrap reduces conservative coverage in the main mixed-frequency comparison. With $T=360$ and $h=6$, coverage is 95.4\% for $N=200$, compared with 99.2\% for the nuisance-covariance plug-in. With $N=1000$, the values are 96.2\% and 97.8\%. The bootstrap mean interval lengths are 0.0146 and 0.0065. Both covariance calculations use the same coefficient estimates and corrections. Positive correlation between neighboring errors can reduce the variance of an exposure contrast with opposite-signed weights, so omitting those covariances can produce conservative intervals.

Mixed-frequency coverage remains close to 95\% in the shorter calendar: at $T=120$ and $h=6$, it is 95.8\% with 200 firms and 95.6\% with 1,000. Short panels can nevertheless exhibit undercoverage. In the block design with $N=1000$ and $T=120$, impact coverage is 88.0\%, and the mean estimated standard error is 0.91 times sampling dispersion. Online Appendix Table~\ref*{tab:localized_patterns} reports the observation-pattern comparisons, and Figure~\ref*{fig:localized_simulation_coverage} gives coverage across horizons.

The bootstrap does not eliminate the coverage shortfall with missing regressors and asymmetric innovations. In DGP~3 with internal gaps, $N=T=240$ and $h=7$, lagged-coefficient coverage is 92.2\%, compared with 91.6\% for the thresholded-score covariance. This comparison retains both the linear and centered quadratic score terms. The improvement is small, and both coverage rates remain below 95\%. Online Appendix Table~\ref*{tab:general_coverage} gives both coefficients at each horizon.

\section{Geopolitical Risk and Firm Investment}\label{sec:application}
We revisit the firm-level investment analysis in \citet[Section~V.B]{CI2022}, who find a larger investment decline in industries more exposed to geopolitical risk (GPR). The quarterly comparison asks how this exposure differential changes as incomplete accounting histories enter estimation. Monthly financial-market information then permits a projection of trailing-three-month investment intensity between fiscal quarter ends. The monthly target and regressor set differ from the quarterly specification, and we interpret their estimates accordingly.

For cross-firm dependence, we use the Standard Industrial Classification (SIC) mapping to the 49 industries of \citet{French49} and text-based network industry classifications (TNIC) from \citet{HP2010,HP2016}. The embeddings-based classification ETNIC provides an alternative peer measure \citep{HP2025}. TNIC-3 and ETNIC-3 identify product-market relationships beyond SIC boundaries. Within each cohort, comparisons of these dependence weights hold the factor fit, exposure measure and analytical correction fixed.

\subsection{Historical replication and quarterly sample}
We begin with the investment specification of \citet[Section~V.B]{CI2022}. Their industry, fiscal-quarter, capital and winsorization rules give a two-quarter FE coefficient of $-0.632$ in the current Compustat vintage, close to the published $-0.630$. The corresponding standard errors are 0.305 and 0.290. Our replication contains 375,497 complete regression cases from 9,264 firms, compared with 374,727 cases in the source table. We retain the published industry exposures and capital-deflator units when extending the accounting and GPR data through 2025. The quarterly covariance estimates use the residual degrees-of-freedom adjustment in Corollary~\ref*{cor:covariance_adjustment}.

The quarterly estimation window is 1995Q1--2025Q4. Eligibility requires at least 60 of its 124 quarters for each accounting variable and the source application's industry and variable-construction restrictions. A further admission screen requires at least 25 jointly observed quarters for every covariance row pair at lags zero through fifteen. This gives 912 firms. Investment, cash relative to capital and Tobin's $Q$ are missing in 12.18\%, 10.45\% and 11.12\% of their firm--quarter cells, respectively. Online Appendix~\ref*{app:sample_construction} describes the admission algorithm and transformations.

Online Appendix Table~\ref*{tab:gpr_missingness} summarizes availability of the underlying quarterly variables. Internal gaps lie between a series' first and last observed values, while leading and trailing missing blocks lie outside that span. Both can occur in the same firm. In the 912-firm cohort, 207 firms have an internal gap in investment, cash relative to capital or Tobin's $Q$. The constructed investment series for Devon Energy has an isolated gap in 2001Q4, and ACI Worldwide has an internal block in 2007Q4--2008Q1. Advance Auto Parts has a leading missing block through 1999Q1 followed by continuous investment observations. These examples describe the availability of the constructed variable, rather than the reason it is unavailable.

\begin{samepage}
The three quarterly samples are named for the histories they admit:
\begin{enumerate}[label=(\roman*)]
\item \emph{Complete quarterly histories}: 232 firms with fully observed investment and accounting controls.
\item \emph{Consecutive histories}: 705 firms, adding 473 whose complete regression cases are consecutive at every reported horizon.
\item \emph{Internal gaps admitted}: 912 firms, adding a further 207 firms with internal gaps.
\end{enumerate}
\end{samepage}
Each expansion retains the firms in the preceding sample. The consecutive-history sample has the history structure used in the MSS unbalanced-panel construction, subject to its common-coefficient and remaining regularity conditions \citep[Appendix~B.2.4]{Mei2026}. FE is computable in each sample. MRLP-BC uses the factor and covariance restrictions to recover the projection over the same period when some of its moments are unavailable.

In the 912-firm sample, 163 firms first report after 1995 and 433 last report before 2025. The full-period quarterly projection extends the investment measure to all included dates. Some source histories include pre-formation periods, so those calculations require an economic interpretation beyond missing reports. We retain the full-period comparison and report the sensitivity analysis using firms with records covering the full period, with 232, 308 and 392 firms, in Online Appendix Figure~\ref*{fig:quarterly_common_calendar}. The main monthly analysis below instead restricts target dates to firm-specific sample periods.

\subsection{Quarterly specification and response paths}
Let $I_{it}$ denote quarterly capital expenditure, $K_{it}$ net property, plant and equipment, and $y_{it}=100\log(100I_{it}/K_{i,t-1})$. Quarterly capital expenditure is obtained from its fiscal-year cumulative series by taking the first-quarter value and subsequent fiscal-quarter differences. Let $g_t$ be the change in log aggregate GPR, standardized over the extended 1985--2025 source sample. The indicator $D_i$ denotes high adverse exposure, constructed from the published industry stock-return sensitivities to GPR using the source paper's firm--quarter median cutoff. Following equation~(5) of \citet{CI2022}, we estimate
\begin{equation}\label{eq:gpr_application}
 y_{i,t+h}=\alpha_{ih}+\beta_h D_i g_t
       +\rho_h y_{i,t-1}+\delta_h g_t
       +\gamma_{h1}c_{it}+\gamma_{h2}q_{it}+u_{it,h},
 \qquad h=0,\ldots,6.
\end{equation}
Cash and short-term investments divided by lagged capital, $c_{it}$, measure liquidity. Tobin's $Q$, denoted by $q_{it}$, is market equity plus total assets less book equity, divided by total assets, and captures investment opportunities. Lagged investment allows for recent investment dynamics, while the separate $g_t$ term permits a response common to both exposure groups. Because the outcome is 100 times log investment intensity, we interpret its changes as percentage changes using the first-order log-change approximation. Following the industry comparison in \citet{CI2022}, $2\beta_h$ measures the difference between the high- and low-exposure groups' percentage investment responses to a two-standard-deviation GPR increase. We report this difference in percentage points. Recovering this conditional projection does not add a source of causal shock identification.

The joint factor model includes investment, the accounting controls, the observed GPR regressors and sales relative to assets. The last variable is complete for 301 firms in the largest quarterly cohort and provides auxiliary information about common variation. All three cohorts use rank three. Transformations, winsorization and scaling are fixed across cohorts, and coefficients are returned to their economic units. The rank comparisons and complete auxiliary counts are reported in Online Appendix~\ref*{app:sample_construction}.

At horizon $h$, base dates run from 1995Q2 through the final quarter whose outcome lies within 2025Q4. The first quarter supplies lagged investment, so $T_h=123-h$ and the largest cohort contains $n=912T_h$ eligible firm--quarter pairs. At $h=2$, 95,413 of 110,352 pairs are complete regression cases. FE uses the complete cases in every cohort. MSS is reported only for complete and consecutive histories, splitting each firm's usable history chronologically into two halves and giving their estimates equal weight. MRLP-BC uses the same observed data and auxiliary series to recover moments for all 110,352 pairs. Within a cohort, the firms and underlying window are held fixed while the estimators use the observed information differently.

For MRLP-BC, we apply the localized multiplier bootstrap to the full joint score, including the factor-fit derivative and restored residual products. The reference weights use TNIC-3 peer profiles and a common 11-quarter Bartlett window for horizons zero through six. The construction and its additional locality conditions are given in Online Appendix~\ref*{app:dated_localization}. Numerical trace covariance is included. Idiosyncratic covariance recovery for the analytical correction retains $L_T=15$ and the 25-pair requirement. FE retains the source application's industry and calendar-quarter clustering, and MSS uses the same clustering on its published score, as specified in Online Appendix~\ref*{app:benchmark_score}. Quarterly intervals are pointwise 90\% intervals, treating membership, rank and classification histories as given.

We report the two-quarter estimates in Table~\ref*{tab:gpr_nested} to maintain comparability with Table~5 of \citet{CI2022}. Their Section~V.D motivates this horizon by the one- to two-quarter delay in the estimated investment response. The coefficient concerns investment two quarters after the GPR change, rather than investment cumulated over those quarters. The two-quarter point estimates change little as firms with incomplete histories enter the sample. For MRLP-BC, the estimated responses to a two-standard-deviation GPR increase are $-0.924$ percentage points in complete quarterly histories, $-0.904$ in consecutive histories and $-0.888$ when internal gaps are admitted. All three pointwise intervals include zero. FE gives negative coefficients of similar magnitude in all three samples, as does MSS on complete or consecutive histories. The MRLP-BC intervals allow for uncertainty in the recovered moments, and the small changes in its point estimates are below the numerical standard errors of the trace corrections in the two incomplete-history samples. These estimates do not resolve a difference between the underlying population responses.

\begin{figure}[htp!]\centering
\caption{MRLP-BC responses across nested firm samples}\label{fig:gpr_nested_paths}
\includegraphics[width=\linewidth]{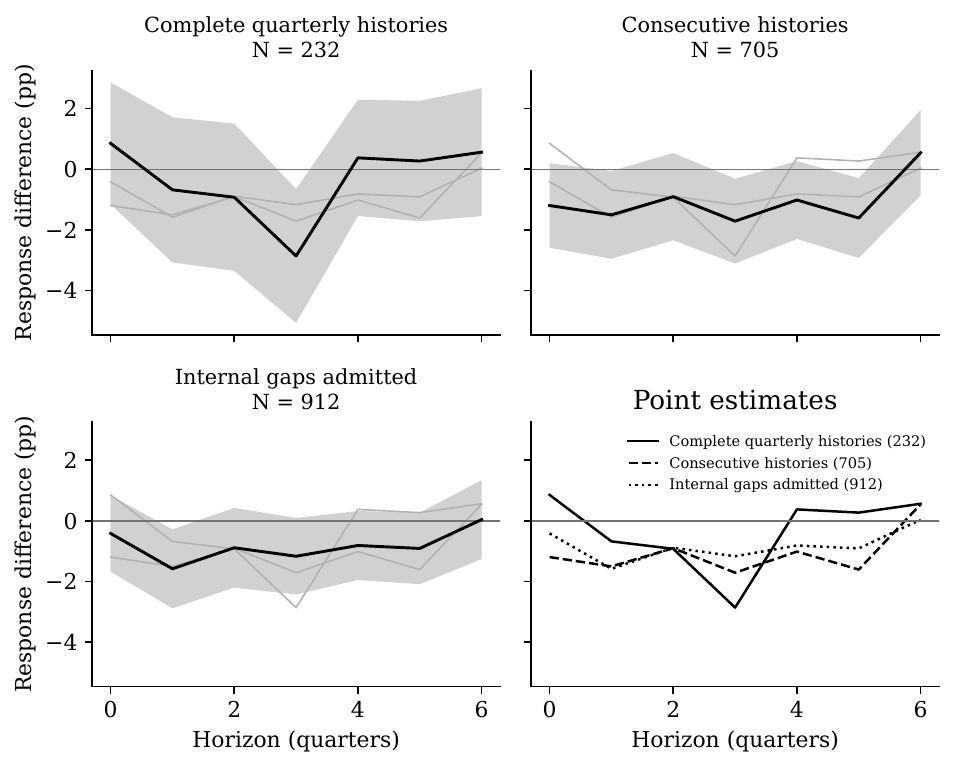}\par\smallskip

\begin{minipage}{\linewidth}\figurenotefont
Notes: Lines show $2\widehat\beta_h$ from equation~\eqref{eq:gpr_application}. These are differences in percentage investment responses between the exposure groups. Three panels show individual sample paths and pointwise 90\% intervals. The fourth compares point estimates. The samples admit complete quarterly histories (232 firms), consecutive histories (705 firms) and internal gaps (912 firms). Selection requires 60 observed quarters per variable and 25 overlapping quarters per covariance pair through lag 15. Membership is fixed across horizons. All three fits use rank three. Intervals use TNIC-3 localized score multipliers with an 11-quarter window and include numerical trace uncertainty. The intervals are neither simultaneous bands nor confidence intervals for differences between samples.
\end{minipage}\end{figure}

The full paths in Figure~\ref{fig:gpr_nested_paths} show differences in timing that the two-quarter comparison does not reveal. Complete quarterly histories have their lowest estimate at quarter three, $-2.861$ percentage points, with TNIC-3 interval $[-5.068,-0.655]$. Consecutive histories reach $-1.711$ at quarter three, with interval $[-3.108,-0.314]$. When internal gaps are admitted, the lowest estimate is $-1.583$ at quarter one, with interval $[-2.882,-0.283]$. ETNIC-3 intervals also lie below zero at these three minima. These marginal results depend on the covariance construction. All FF49 pointwise intervals include zero, and the 90\% simultaneous bands under either peer construction contain the zero path in every quarterly sample. The estimated timing differences remain descriptive because firm composition, the estimated factor space and projection moments change together. Online Appendix Figure~\ref*{fig:quarterly_common_calendar} gives the quarterly sensitivity analysis for the 232, 308 and 392 firms whose records cover the full period.

\subsection{Monthly investment projections for four nested firm samples}\label{subsec:monthly_application}
For fiscal quarter $q$ ending in month $t$, the observed outcome is $Y_{it}=100\log(100I_{iq}/K_{i,q-1})$, using that quarter's capital expenditure and capital at the preceding fiscal quarter end. We date it by Compustat's fiscal-period-end date and recover the conditional mean of the corresponding trailing-three-month intensity between observations. Monthly CRSP returns and log market capitalization from the same 411 firms, all fully observed, provide auxiliary information in every sample. Annual fundamentals remain annual measurements. This is a retrospective projection of a trailing-three-month measure, rather than a disaggregation into individual monthly investment flows. Online Appendix~\ref*{app:sample_construction} gives the joint-fit scaling.

For each firm, the main monthly analysis uses the months between its first and last quarterly records within the study period. We call this its firm-specific sample period. Both the month of the GPR change and the outcome month must fall within this period. Within-unit means, projection weights, normalization and the analytical correction are computed over these dates, using Proposition~\ref*{prop:known_windows} and Corollary~\ref*{cor:window_lmb}. The four nested samples contain 232, 705, 912 and 1,174 firms. The last sample meets the requirement of 60 observed quarters for each accounting variable, without the quarterly restriction on overlap across all covariance pairs. At monthly frequency, nuisance covariance estimates based on fewer than 25 joint observations are set to zero, while the firms remain in the sample. Table~\ref{tab:monthly_missing} separates missing observations within each firm's sample period from dates outside it.

\begin{table}[htp!]\centering\small
\caption{Investment-data availability by sample period, 1995--2025}\label{tab:monthly_missing}
\setlength{\tabcolsep}{2pt}
\begin{tabular*}{\linewidth}{@{\extracolsep{\fill}}lrrrr@{}}\toprule
 & \shortstack{Complete\\quarterly histories} & \shortstack{Consecutive\\histories} & \shortstack{Internal gaps\\admitted} & \shortstack{All eligible\\firms}\\\midrule
Firms & 232 & 705 & 912 & 1,174\\
Firm-months in the full period & 86,304 & 262,260 & 339,264 & 436,728\\
Within each firm's sample period & 86,289 & 239,507 & 310,223 & 370,186\\
Observed investment entries & 28,768 & 77,357 & 99,382 & 118,181\\
Missing within the sample period & 57,521 & 162,150 & 210,841 & 252,005\\
Outside the sample period & 15 & 22,753 & 29,041 & 66,542\\
Minimum observed reports per firm & 124 & 62 & 60 & 60\\
Median observed reports per firm & 124 & 120 & 117 & 102\\
Distinct observation patterns & 2 & 204 & 374 & 516\\
\bottomrule\end{tabular*}\par\smallskip
\begin{minipage}{\linewidth}\tablenotefont Notes: The calendar contains 372 months. Observed investment entries are dated by their fiscal quarter ends. Unobserved months within firm-specific sample periods include routine gaps between quarterly measurements and additional unavailable observations. Dates outside each firm's sample period do not enter the main projection. The eligible regression dates also depend on the horizon, so these counts are not the number of observations used at every horizon. Observed values, missing values within each sample period and dates outside it sum to the total for the full period. Cohort names refer to quarterly accounting histories.
\end{minipage}\end{table}

Let $g_t^M=\Delta\log(\mathrm{GPR}_t)$ and let $D_i$ be the source paper's high-exposure indicator. The monthly specification is
\begin{equation}\label{eq:monthly_application}
\begin{split}
 Y_{i,t+h}&=\alpha_{ih}^M+\sum_{j=0}^3\beta_{hj}^M D_i g_{t-j}^M
                +\sum_{j=0}^3\delta_{hj}^M g_{t-j}^M\\
 &\quad+\gamma_h^{M\prime}m_{t-1}+u_{it,h}^M,
 \qquad h=0,\ldots,18.
\end{split}
\end{equation}
Here $m_{t-1}$ contains lagged industrial-production growth, PCE inflation and the federal funds rate. Growth and inflation are 100 times monthly log changes in INDPRO and PCEPI. All eleven regressors are observed. The specification omits monthly lagged investment and the quarterly accounting controls. The monthly GPR change is unstandardized. A 10\% GPR increase gives the high-minus-low response differential $\log(1.10)\beta_{h0}^M$, in percentage points. We estimate the conditional projection of Section~\ref{subsec:conditional}, maintaining mean independence of the outcome error from the entire regressor path and the recorded windows.

TNIC-3 multiplier weights use peer profiles that vary over time. We use a six-month Bartlett window and the joint numerical trace covariance. Within a fixed 1,174-firm reference universe, each firm selects up to eight mutual peers. The normalized indicator of the firm and these peers gives the feature vector in Proposition~\ref*{prop:peer_features}. Profiles are updated each July using prior-year information. The local TNIC-3 and ETNIC-3 files end in 2021 and 2023, respectively, and the last available profiles are carried forward. Unmatched firms remain in estimation. Online Appendix~\ref*{app:industry_data} documents coverage and classification timing.

\begin{figure}[htp!]\centering
\caption{Monthly investment responses over firm-specific sample periods}\label{fig:monthly_four}
\includegraphics[width=\linewidth]{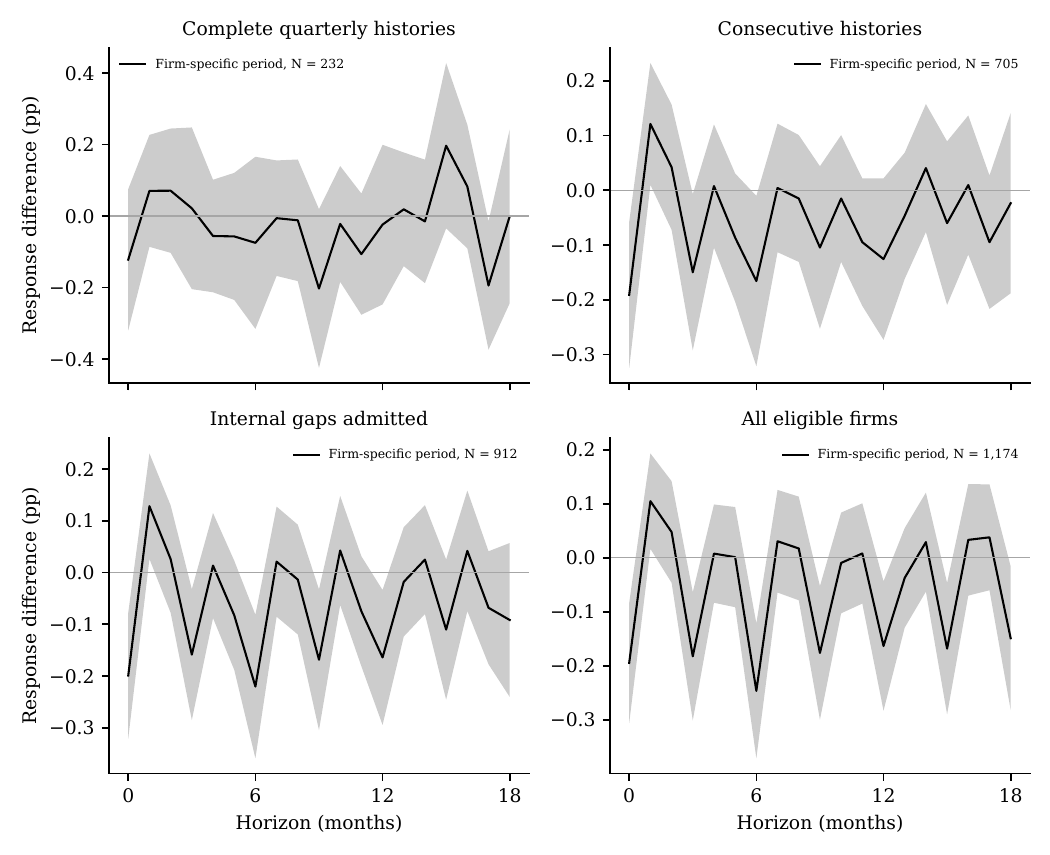}\par\smallskip
\begin{minipage}{\linewidth}\figurenotefont Notes: Responses are exposure-group differentials following a 10\% GPR increase, in percentage points. The regression and outcome months lie inside the period between each firm's first and last quarterly records. Shading gives nominal 95\% TNIC-3 localized multiplier bootstrap pointwise intervals. All horizons use the same 411 auxiliary firms and rank three. Within-unit means, analytical corrections and numerical trace covariance are computed for the stated sample periods. Cohort names refer to quarterly histories. Intervals concern each stated population and are not tests for differences between curves.\end{minipage}\end{figure}

Figure~\ref{fig:monthly_four} reports the response paths over firm-specific sample periods. At six months, the response differences between high- and low-exposure firms following a 10\% GPR increase are $-0.075$, $-0.166$, $-0.220$ and $-0.246$ percentage points for the four cohorts. The 1,174-firm estimate has nominal 95\% pointwise interval $[-0.371,-0.121]$. The six-month differential becomes more negative as firms with incomplete histories enter the analysis. Across cohorts, both the firms and their projection dates change, so these intervals concern each stated projection and do not test equality of the curves.

The paths alternate within three-month cycles, with recurring declines at quarter-spaced horizons and smaller responses at intervening months. The six-month comparison is therefore one part of the path. The outcome remains trailing-three-month investment intensity, so this pattern does not by itself identify the timing of individual monthly expenditures. Simultaneous bands for all nineteen horizons are in Online Appendix Figure~\ref*{fig:industry_simultaneous}. ETNIC-3 gives similar six-month uncertainty. FF49 weights provide a sensitivity analysis based on industry dependence, with the point estimates fixed. Their fixed number of large groups lies outside the sufficient condition on the localization rate in Theorem~\ref{prop:dated_covariance}.

We also estimate the response over the common period 1995--2025 in two samples. Using every firm gives a six-month estimate of $-0.321$ percentage points for 1,174 firms. Restricting the sample to firms with records covering both ends of the period retains 232, 308, 392 and 395 firms, with an estimate of $-0.266$ in the largest sample. Online Appendix~\ref*{app:target_comparison} gives both response paths and their intervals. Using every firm requires extending the investment measure beyond its recorded history, including dates before some firms were formed. Requiring records at both ends avoids that extrapolation but still requires an economically meaningful outcome and a consistent accounting entity throughout the period.

The observed data must contain enough factor variation to estimate each firm's loadings. The smallest eigenvalue of the observed factor Gram matrix is 0.253 in the complete-history sample and 0.000791 in the sample of all eligible firms. Calibration using the observed missing-data patterns gives six-month coverage of 94.4\% and 91.6\% for the 232- and 1,174-firm cases with peer dependence, and simultaneous coverage of 96.8\% and 89.8\%. The procedure is re-estimated in every replication, but the calibration uses the full sample period and time-varying peer covariances. It therefore does not establish coverage for the firm-specific sample periods or verify the locality condition. Online Appendix~\ref*{app:application_calibration} describes the design and the case with strong residual dependence.

As a sensitivity check, we require observed histories to retain at least one-tenth of the factor variation in every direction, measured using the reference factors from the auxiliary data. The cutoff is 0.10, and the same reference basis is used in each sample. This restriction leaves 232, 438 and 572 distinct firms. Using firm-specific sample periods, the 572-firm six-month estimate is $-0.106$ percentage points, with interval $[-0.263,0.050]$. The smaller magnitude accompanies a change in both the firms and the estimated factor model. The cutoff is a numerical choice for this sensitivity check. The nominal intervals do not account for selection using estimated factors. Online Appendix~\ref*{app:monthly_screen} gives the normalization and the complete response paths.

\section{Conclusions}\label{sec:conclusions}
Missing data can remove every complete regression case while leaving the projection normal equations identified. Strong auxiliary information recovers the common-component products, and observed residual pairs recover the idiosyncratic contribution. MRLP-BC corrects the leading within-transformation and fitting biases before inverting those moments. The complete score then supports inference by the localized multiplier bootstrap under the stated dependence restrictions. With fully observed regressors, the same argument yields a linear score and permits prespecified unit-specific projection windows.

The investment application revisits \citet{CI2022} using incomplete quarterly histories and monthly financial-market data. Over firm-specific sample periods, the six-month response difference between high- and low-exposure firms is more negative in the larger samples. Estimates over the common sample period are also negative in the larger samples. Restricting the sample by observed factor variation reduces their magnitude. These comparisons concern different firms and estimation periods. Their interpretation relies on the stated factor and dependence assumptions, and the calibration shows that finite-sample coverage remains a concern.

\FloatBarrier
\bibliographystyle{ecta}
\bibliography{references}

\end{document}